\documentclass[useAMS,usenatbib]{mn2e}

\usepackage{epsfig}

\newcommand{\kms}{km~s$^{-1}$}

\title{Cosmic Flows:  Green Bank and Parkes HI observations }

\author[Courtois et al.]
{
H\'el\`ene M. Courtois$^{1,2}$\thanks{E-mail: \texttt{h.courtois@ipnl.in2p3.fr}}, 
R. Brent Tully$^{2}$, 
D. I. Makarov$^{1,3}$, 
S. Mitronova$^{3}$, \newauthor
B. Koribalski$^{4}$,
I. D. Karachentsev$^{1,3}$
and J. Richard Fisher$^{5}$\\
$^{1}$Universit\'e Lyon 1, CNRS/IN2P3/INSU, Institut de Physique Nucl\'eaire, Lyon, France\\
$^{2}$Institute for Astronomy, University of Hawaii, 2680 Woodlawn Drive, Honolulu, HI 96822, USA\\
$^{3}$Special Astrophysical Observatory, Russian Academy of Sciences, N. Arkhyz, KChR, 369167, Russia\\
$^{4}$Australian Telescope National facility, CSIRO,  PO Box 76, Epping NSW 1710, AUSTRALIA\\
$^{5}$National Radio Astronomy Observatory\thanks{The National Radio Astronomy Observatory is a facility of the National Science Foundation, operated under cooperative agreement by Associated Universities, Inc.}, 520 Edgemont Road, Charlottesville, VA 22903, USA
}

\begin{document}

\date{Accepted xxx. Received xxxx; in original form 2010 December 1}

\pagerange{\pageref{firstpage}--\pageref{lastpage}} \pubyear{2011}

\maketitle

\label{firstpage}

\begin{abstract}
The neutral hydrogen properties of 1,822 galaxies are being studied with the Green Bank 100m and the Parkes 64m telescopes as part of the `Cosmic Flows' program.  Observed parameters include systemic velocities, profile line widths, and integrated fluxes.  The line width information can be combined with optical and infrared photometry to obtain distances.  The 1,822 HI observations complement an inventory of archives.  All told, HI line width information is available for almost all of five samples: (i) luminosity--line width correlation calibrators, (ii) zero-point calibrators for the supernova Ia scale, (iii) a dense local sample of spiral galaxies with $M_{Ks} < -21$ within 3,000~\kms, (iv) a sparser sample of 60~$\mu$m selected galaxies within 6,000~\kms\ that provides all-sky coverage of our extended supercluster complex, and (v) an even sparser sample of flat galaxies, extreme edge-on spirals, extending in a volume out to 12,000~\kms.  The HI information for 13,941 galaxies, whether from the archives or acquired as part of the Cosmic Flows observational program, is uniformly re-measured and made available through the Extragalactic Distance Database web site.
\end{abstract}

\begin{keywords}
astronomical data base; catalogs; galaxies: distances; radio lines: galaxies
\end{keywords}

\section{Introduction}

Cosmography is the study of the large scale structure of the universe.  A complete analysis involves observational and interpretive components.  With spectroscopic information, a complete sample of galaxies within specified limits describes structure in redshift space.  If distance measures are available for at least some fraction of the spectroscopic sample then steps can be taken to transform to physical space.  Distance measures allow separation of redshifts into cosmic expansion and deviant (or peculiar) components.  Cosmological simulations and orbit reconstructions provide tools for the recovery of information about the underlying distribution of matter from a map of peculiar velocities.  The influence of the dark sector on galaxy motions can be studied from $\sim 1$~Mpc, the scale of collapse, to $\sim 150$~Mpc, the largest scale of useful peculiar velocity measures.  

After great enthusiasm for what could be learned from peculiar velocity studies in the 1990's \citep{1998ApJ...507...64W}, \citep{2000ASPC..201...17C}  progress slowed primarily because of the challenge presented by the need for much more and much better data.  If the goal is to have a dense grid of distance measures to a depth dominated by Hubble expansion then most methodologies for estimating distances have inadequacies.  The Cepheid Period-Luminosity \citep{2001ApJ...553...47F}  and Tip of the Red Giant Branch \citep{2007ApJ...661..815R} methods have limited reaches.  The Surface Brightness Fluctuation \citep{2001ApJ...546..681T} and Fundamental Plane \citep{2001MNRAS.321..277C} methods apply to luminous early type galaxies that are poorly represented in low density regions.  The Type Ia Supernova \citep{2007ApJ...659..122J}  method rests on serendipity, resulting in an accurate but sparse map of distances.  The one well established methodology that can provide decent distances with high density over an appropriately large volume is  provided by the correlation between galaxy luminosity and rotation rate, the Tully-Fisher Relation \citep{1977A&A....54..661T}.  

Two observations are needed to apply this method: a spectroscopic measure of the rotation rate, most expeditiously accomplished by observing the line width of a 21 cm Neutral Hydrogen profile, and surface photometry at optical or infrared bands that monitor old star populations.  New capabilities with both spectroscopy and photometry are revolutionizing our capabilities.  On the spectroscopic side, the new capabilities are both realized and promised.  We will describe observations with the 100m Robert C Byrd Green Bank Telescope at the National Radio Astronomy Observatory (NRAO-GBT) and with the  13 channel Multibeam Receiver on the 64m Parkes Telescope.  Our program also makes use of archival data from ongoing multibeam observations with the Arecibo Telescope (Giovanelli et al.  
2005).  The promised capabilities are those that will accrue with the wide-field interferometric surveys that will cover the entire southern (ASKAP, MeerKAT) and northern hemispheres (Apertif, LOFAR) with unprecedented sensitivities.  As for the photometry, the ground technology revolution comes from the large multi-band surveys from new  wide-field CCD camera systems
Pan-STARRS in the north and SKYMAPPER in the south. Also the satellites Spitzer and WISE are providing unprecedentedly accurate surface photometry of galaxies in the mid-infrared.

The data must be compared with analytical or numerical models. The theoretical tools have developed tremendously in the last 20 years from small N-body programs to extremely large hydrodynamical simulations that have culminated with such data products as the Millennium Simulation
Project \citep{2005Natur.435..629S}.  The advances have focused on increasing computational speed, the number of particles, and parallelism in the codes. The field is now embracing higher levels of refinement in the constraints on initial conditions, coming from the observations.
The theoretical universe being built in the new generation of simulations must implement  observational evidence with a high level of detail. In particular on scales of 1-150 Mpc, there is a tension between the observed luminous matter distribution and the complexity of galaxy flows, both in directions and in amplitudes that must be reconciled within theoretical constructs.  New developments in analysis methods \citep{1995ApJ...449..446Z} use information in the gradients and convergence of flows to
recover mass distribution information that extends beyond the observational data.  There is an encouraging synergy in the advances on both observational and theoretical fronts.
 
This conjunction of observational and theoretical progress creates a special opportunity for the emergence of the `Cosmic Flows' program that was proposed to NRAO as a large program of observations of the HI 21cm line of galaxies with the GBT telescope.
 It is ongoing since 2007 and already has been scheduled for more than 1,000 hrs on the sky. The radio observational program is extended to the full sky with access to the most southern targets using the ATNF-CSIRO Parkes 64m radiotelescope. An accompanying photometry program is designed to provide the surface photometry of
 the targets. The theoretical program accompanying the observational program  is using Numerical Action Methods \citep{2001ApJ...554..104P} and numerical constrained simulations: CLUES \citep{2010arXiv1005.2687G} .
 
 The present paper will focus on 1) a presentation of the five complementary data samples of the Cosmic Flows project, 2) a description of the pipeline developed for the consistent measurement of tens of thousands of HI profiles (in anticipation of the coming large surveys), and 3) the release of the  currently accrued radio HI material associated with our five samples.

\section{ Cosmic Flows Program: Five galaxy samples }

The goal of the Cosmic Flows program is to obtain an all-sky grid of galaxy distances as dense and deep as current capabilities allow and to complement the observational endeavor with collaborative theoretical studies to try to gain a more rigorous understanding of the local dark sector (http://www.ifa.hawaii.edu/cosmicflows/). The volume of universe that is currently targeted is bounded by the sensitivity of the GBT and the Parkes single dish radio telescopes at a practical limit of 6,000 \kms\ radius.  It is expected that the next generation of the Cosmic Flows program will be testing a volume up to 15,000 \kms\ using wide field interferometric technology with blind HI all-sky surveys.  For the moment, competition for telescope time makes it impractical to target even the $\sim 7000$ appropriately edge-on spiral systems with measured redshifts within 6,000 \kms.  Our compromise has been to distinguish three discrete samples and to strive for a high level of completion with each of the three components.
One sample, we refer to as "V3K", strives to provide a high density mapping of the volume within 3000~\kms.  A second sample selected by far infrared flux, our "PSCz", extends to 6000~\kms\ and gives good coverage at low Galactic latitudes.  The third sample of extreme edge-on systems, "RFGC", provides sparse coverage over a large volume.
Two smaller samples containing several hundred targets were observed for calibration purpose.  One provides the slope and zero-point calibrations for the Tully-Fisher (TF) relation.  The other provides a scale coupling with the Supernova of Type Ia (SNIa) method.  These five samples are described in more detail below.

While our new observations were focused on the five samples to be described, our archival searches have been unrestricted.  We have gathered all available digital HI spectra and coherently remeasured the HI profiles for 16,004 spectra of 13,941 galaxies, for which 11,074 galaxies have adequate measurement to derive distances.  Our observations were thus targeting only the galaxies within our specified samples and without adequate HI profile measurements.  These limitations have allowed us to essentially complete coverage of our samples with about 1,200 observed targets in 3 years.
As points of comparison, in recent discussions based on literature data, Tully et al. (2008)  is based on 1791 distance within 3,000 \kms\ including 1252 TF measures, and Feldman et al. (2010) is based on 4536 distances dominated by the TF measures of \citep{2005ApJS..160..149S}.

Our goal was not merely to incrementally augment the literature information.  The current procedures involve a new definition of the HI linewidth and, as a consequence, required new measurements of 100\% of the profiles and a complete re-calibration of the Tully-Fisher relation.  

The number of calibrators have increased by more than 100\% since the previous extensive calibration of \citep{2000ApJ...533..744T}, giving an additional good reason to reconsider the calibration.  The new procedures and calibration will then be applied to the samples of field galaxies.  We measure
the integrated flux of the HI linewidth, and then derive the profile width at 50\% of this cumulative flux. This method requires access to electronic profiles but then enables re-computation of all the available profiles in all archives.  Our procedures have been compared with similar procedures by the Cornell group  and with an accurate non-digital sample  \citep{2009AJ....138.1938C} with satisfactory results.

\subsection{Calibrators sample}
The luminosity--linewidth calibration divides into two parts;
the first related to the slope of the correlation and the second with the determination of the zero--point. 

The determination of the slope of the
luminosity--linewidth relation is critical for the minimization of systematic
errors due to one of the two Malmquist biases; the effect on errors that can result from magnitude-limited samples (Malmquist 1924; \citep{1984A&A...141..407T} ; \citep{1994ApJS...92....1W} ; \citep{1994ApJ...430...13S} ) \footnote{There is confusion because Malmquist (1920, 1922, 1924) discussed possible biases that affect galaxy distance measurements in two distinct ways.  Suppose individual galaxy distances are unbiased but have errors.  Since there are an increasing number of sources at larger distances, a given shell in measured distance will be populated by more galaxies with larger true distances scattered inward than smaller true distances scattered outward.  Lynden-Bell et al. (1988) drew attention to this effect in the context of galaxy distances and flows and subsequently these authors have referred to it as the Eddington-Malmquist bias in recognition of a prior contribution by Eddington (1913).  The other relevant effect is caused by a magnitude limit.  For galaxies at the same distance, intrinsically fainter systems can be lost from a sample while intrinsically brighter systems are retained.  \citep{1995PhR...261..271S}, in their extensive discussion of the subject, refer to this latter effect as a `selection bias' and identify `Malmquist bias' with the Eddington-Malmquist effect. In this paper, we will continue to use the term Malmquist bias in connection with the magnitude selection problem because of the past popular usage and because for what we are doing it tends to be the greater concern.}.
A maximum likelihood procedure such as that employed by \citet{1997AJ....113...22G} can be used that provides a basis for corrections for bias if there is an accurate understanding of sources of errors.  Instead, our procedure, discussed by  
\citet{2000ApJ...533..744T}, does not require a detailed understanding of errors to correct biases but, rather, acts to "null" the bias.  Individual distance measures have uncertainties but are not systematically offset from true distances.

An essential ingredient to the recipe to null bias is a slope fit to the luminosity--linewidth correlation that is insensitive to magnitude limits.   The desired calibration is achieved through the  compilation of a cluster template created by successively combining data from individual clusters with offsets in magnitude reflecting their different moduli.   The calibration is achieved with consideration of  370 galaxies in 13 clusters of diverse morphologies with distances ranging from 15 to 100 Mpc.   Each cluster contribute 15-70 galaxies (median of 25) to the template.
Each cluster sample is magnitude limited and not otherwise limited in a parameter that constrains the fit for a derivation of distances \footnote{Galaxies with morphological types Sa and earlier are excluded.  This selection has a qualitative aspect that could introduce a small bias.  Also, potential candidates are excluded if there is confusion in the radio telescope FWHM beam or strong evidence of tidal disruption.  Galaxies are excluded if they are more face-on than $45^{\circ}$ but this property is not intrinsically correlated with distance and tests have not suggested any distance bias with our inferred inclinations.}.

The absolute zero point calibration of the template relation is provided by
39 galaxies that pass the same inclination, type, and luminosity criteria as
the cluster calibrators and have accurately known distances from external
measurements.  These calibration distances are based on Hubble Space Telescope observations of either Cepheid stars or the luminosities of stars at the Tip of the Red Giant Branch,
with the scale set by the HST distance scale key project \citep{2001ApJ...553...47F}.  
In \citet{2008ApJ...676..184T} we showed a preliminary result demonstrating that TRGB, SBF, and luminosity--line width (TF)  distances are all on a 
consistent zero point scale with those established by Cepheid measurements.
The previous zero point calibration in this collaboration was based on 24 galaxies \citep{2000ApJ...533..744T}, so a new calibration is possible using 60\% more galaxies.

To complete these calibrator samples, we observed an additional 165 galaxies with no previous
satisfactory measurement with the NRAO-GBT Large Program.
We now have digital data for the entire sample. The linewidth and flux 
analysis that was described in \citet{2009AJ....138.1938C} was  performed in a consistent way.  The cluster slope calibration sample
with accurate line widths has reached 326 galaxies, an improvement of more than 100\% compared to the 155 galaxies in 12 clusters in  \citet{2000ApJ...533..744T}.

\subsection{SNIa host galaxy sample}

The SNIa host galaxy sample is designed to provide a  confident link between the local
scales probed by luminosity--line width distances and dominated by 
peculiar motions and the cosmological scales probed by SNIa distances.
To built this sample we extracted from the literature spiral edge-on galaxies suitable for the TF relation and that have hosted a SNIa. The SNIa must have a well sampled light-curve and an accurate distance measurement. We draw samples from  \citet{2007ApJ...659..122J} and \citet{2003ApJ...594....1T}.  With the GBT observations program we completed the HI observations
of a sample of 54 galaxies which should provide a robust
inter--calibration of the luminosity--line width and SNIa distance 
scales.

\subsection{V3K sample}

A special focus on the region within $V_{helio} = 3,300$ \kms\ derives from three considerations.  First of all, this velocity bounds the structure that historically has been called the Local Supercluster (de Vaucouleurs 1953) with the Virgo Cluster at its core and the Fornax Cluster as a secondary feature well within the boundary.  Secondly, with velocity measures of 15\% accuracy, peculiar velocities can be separated from expansion velocities with uncertainties of less than a few hundred \kms\ for individual objects and errors in collective motions of relatively small groups of galaxies can be brought below 100 \kms.  Third, essentially all galaxies that might be typed Sb to Sd can be easily detected in HI at high signal-to-noise ratio anywhere in the sky if they lie within 3,300 \kms.

Our rigorous V3K sample is defined by the following criteria:
 
$-$ $V_{helio} < 3,300$ \kms
 
$-$ $M_{Ks} < -21$

$-$ inclination from the axial ratio of greater than $45^{\circ}$

$-$ type later than Sa

$-$ no pronounce evidence of tidal disruption

$-$ HI signal not confused from a second galaxy 

The clip in magnitude at $M_{Ks}=-21$ minimizes selection problems.  
The $Ks$ (short $K$) magnitudes are drawn from the 2MASS Extended Galaxy Catalog and Large Galaxy Atlas \citep{2003AJ....125..525J} assuring uniform coverage of the sky relatively unaffected by obscuration down to low Galactic latitudes.  Absolute magnitudes are derived from velocities
and a non-parametric model of galaxy motions \citep{1995ApJ...454...15S}.
Since the faintest galaxies have $Ks=12$ and are gas-rich spirals the 
sample is quite complete at high latitudes and all candidates are easily 
detectable in HI with GBT.

\begin{figure*}
\begin{tabular}{cc}
\includegraphics[width=8cm]{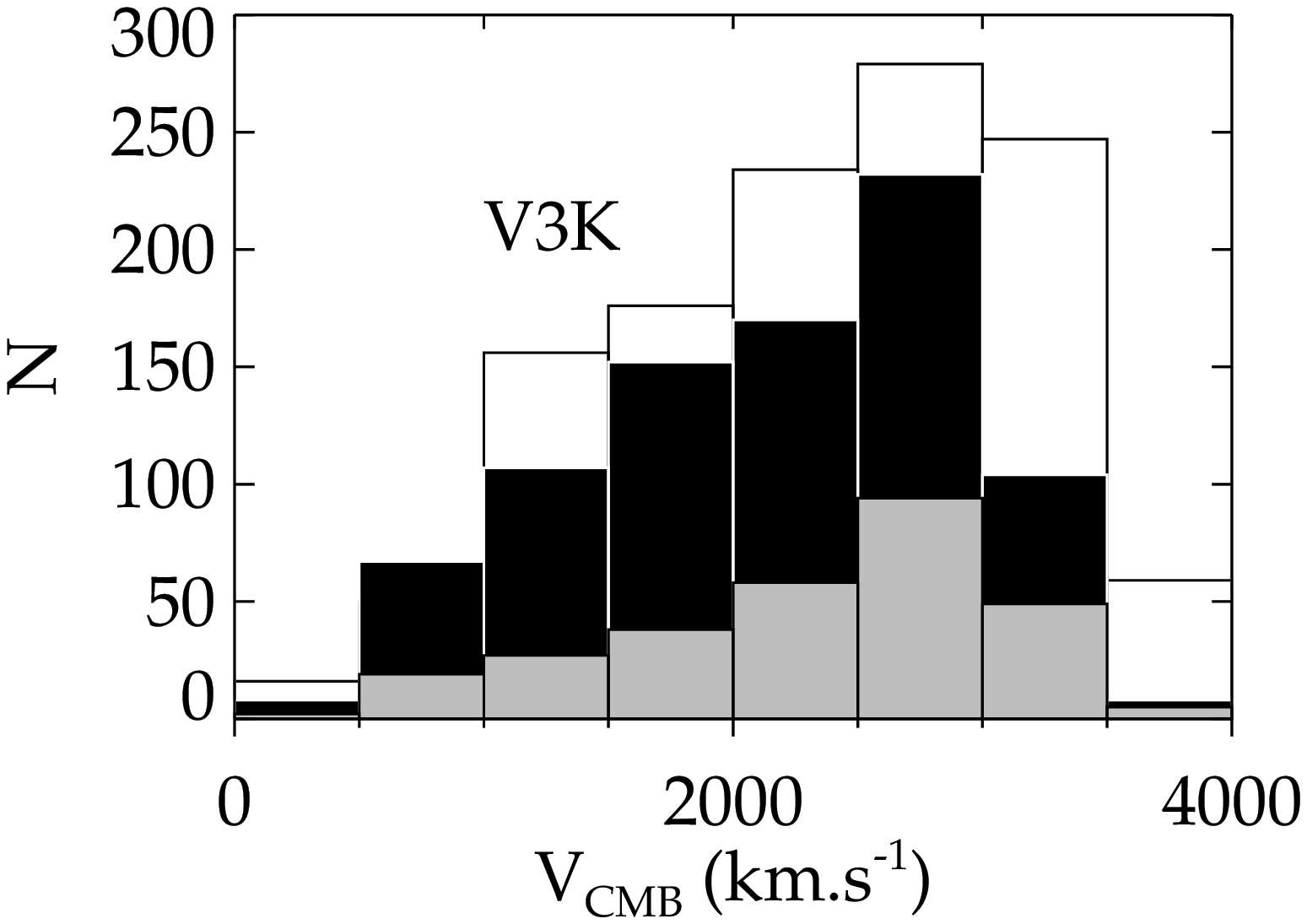} &
\includegraphics[width=8cm]{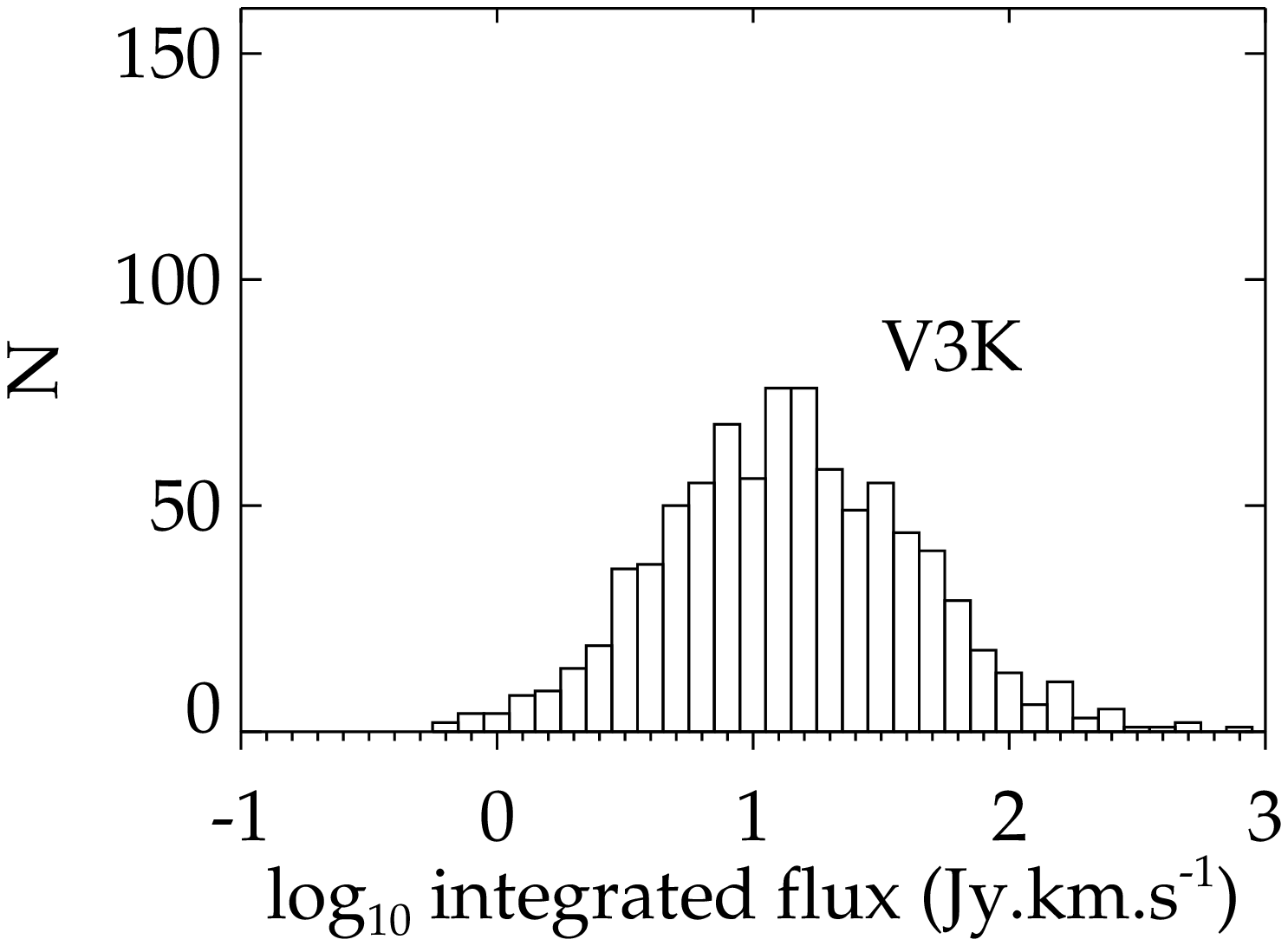} \\
\includegraphics[width=8cm]{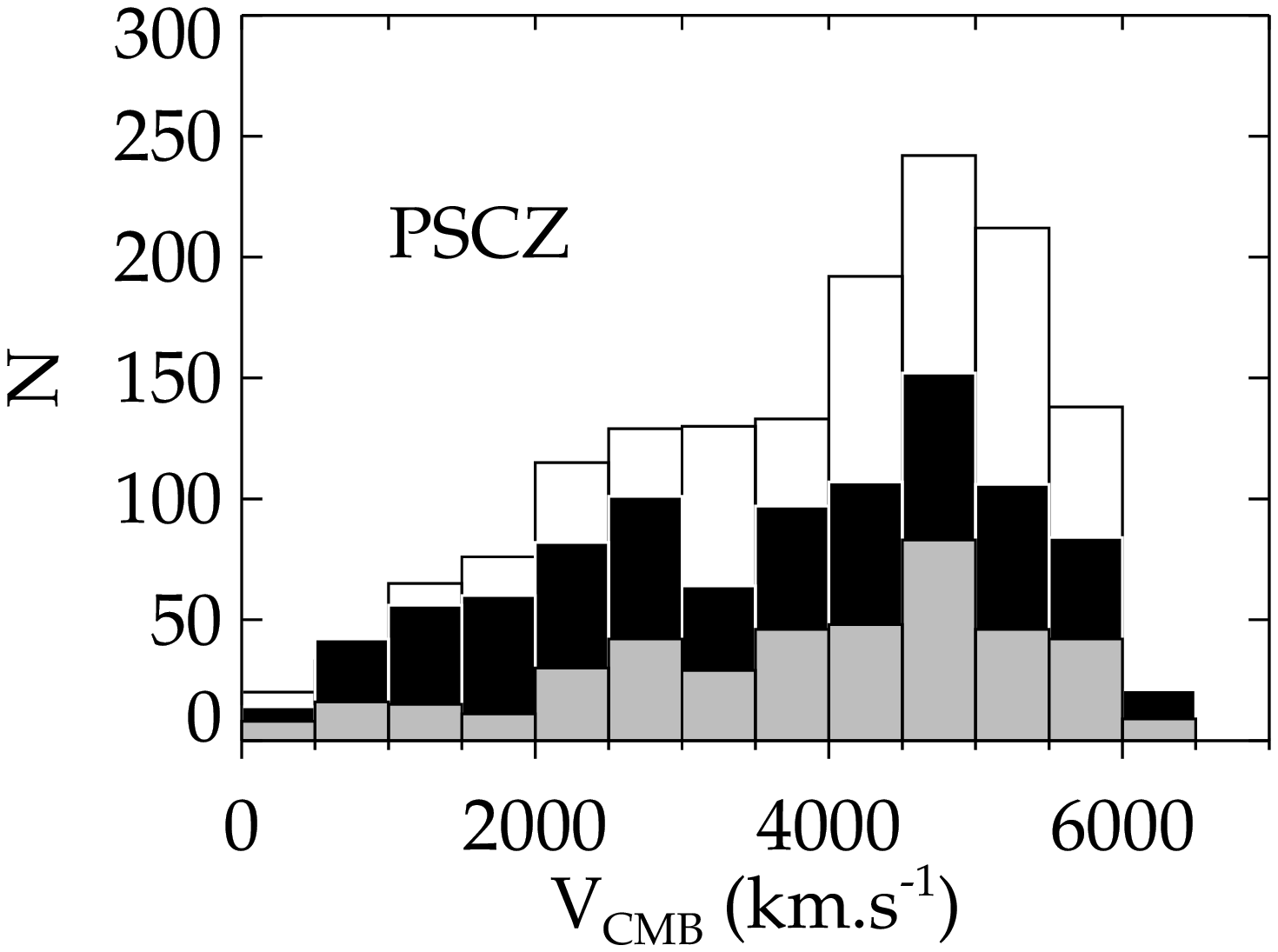} &
\includegraphics[width=8cm]{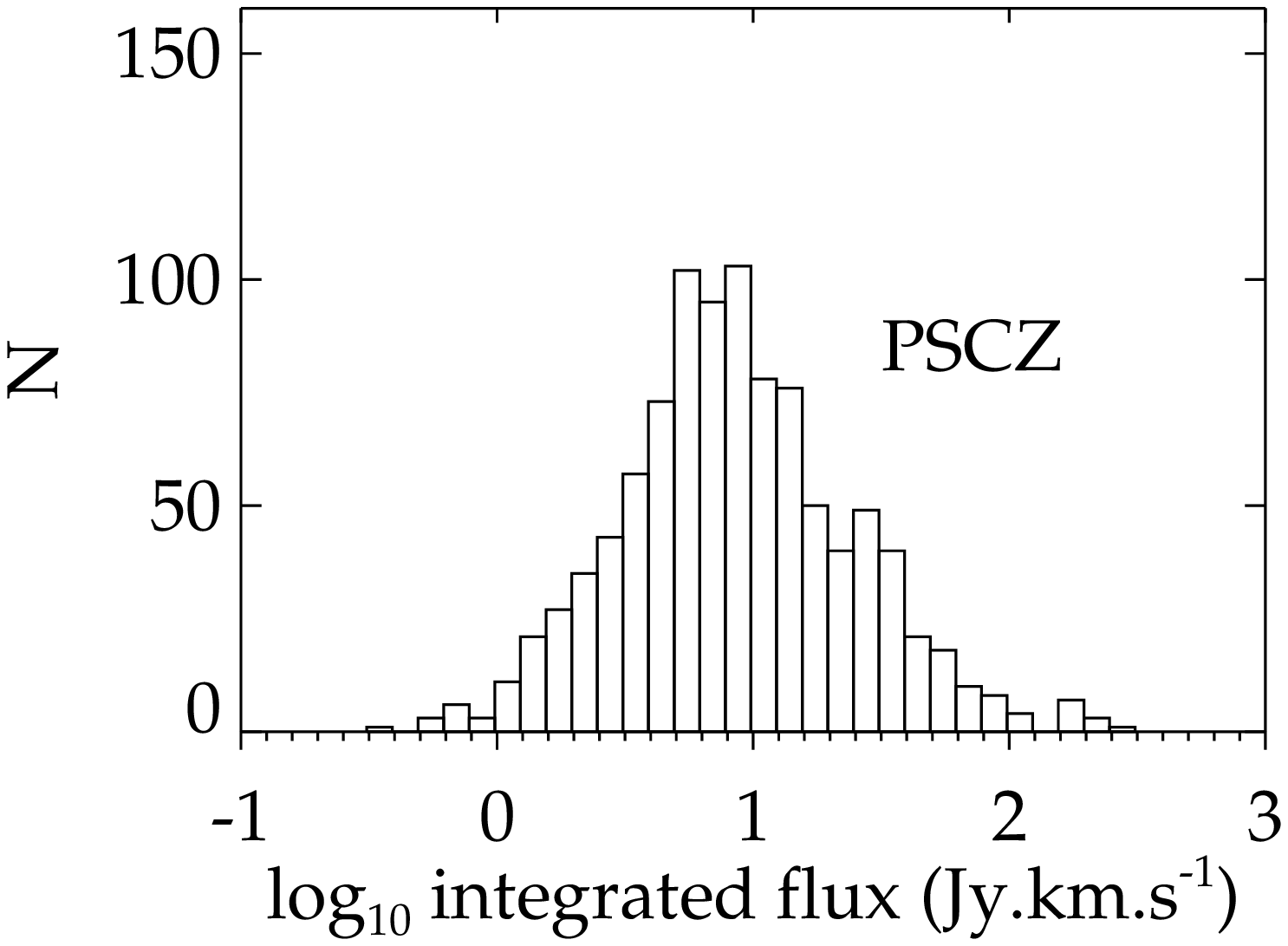} \\
\includegraphics[width=8cm]{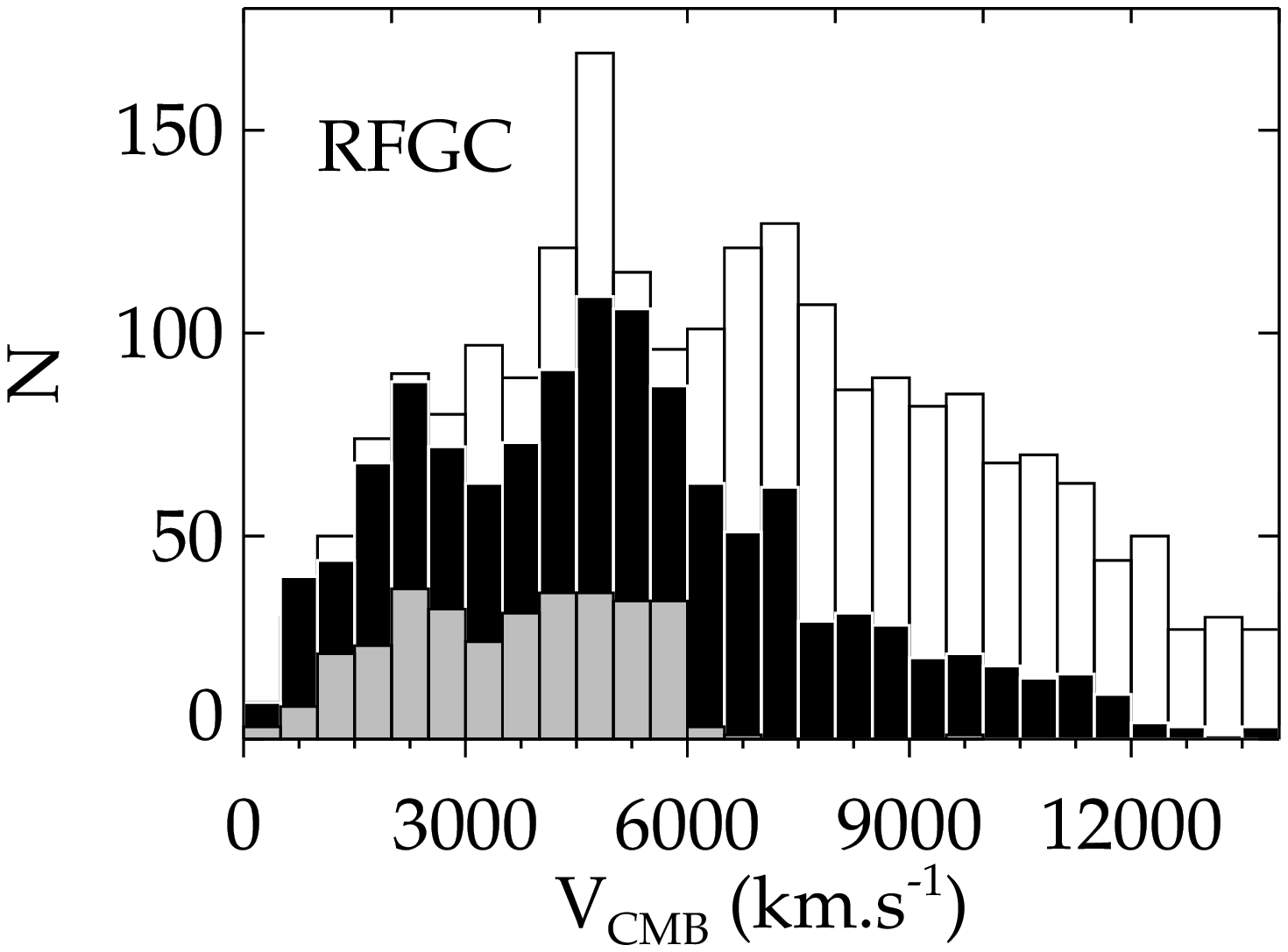} &
\includegraphics[width=8cm]{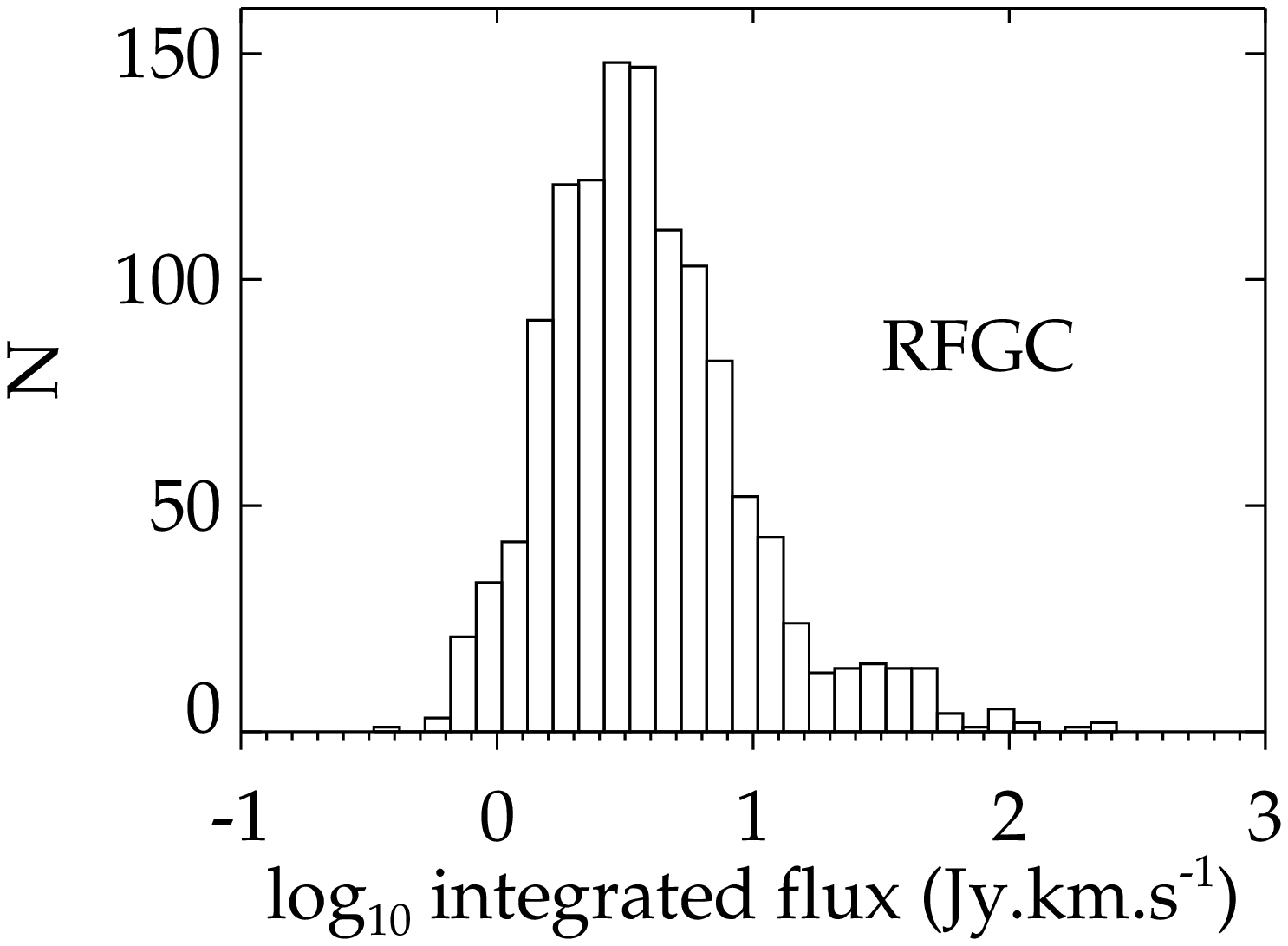} 
\end{tabular}
\caption{
The original Cosmic Flows samples from top to bottom:  V3K, PSCz and Flat Galaxies  (white histograms) compared to  the total of accurate linewidth measurements we have measured (black histograms) and to our GBT and Parkes pointed observations (grey histograms). All our samples are now completed in HI observations apart from the region of the sky better observed with Arecibo and Parkes. Some galaxies were dropped from the original sample usually because of beam confusion or inadequate inclination. In the right panels, histograms illustrate the distributions of source integrated fluxes in HI.  The galaxies in the PSCz and RFGC samples are progressively more distant on average than the V3k galaxies so are progressively fainter in apparent flux. 
}
\label{3samples}
\end{figure*}

A velocity histogram of the 1,228 galaxies in V3K is shown as seen in Figure~\ref{3samples}.  
The original V3K sample contained 1,228 galaxies.  Of these,
993 galaxies have now measured HI spectra of adequate quality and, of these,
852 have consistently measured digital profiles.
The Cosmic Flows observations provided 292 new accurate measurements while 560 accurate measurements could be recovered
from the re-analysis of archive material.
The 135 galaxies that were dropped from the initial sample frequently have a companion galaxy in the area included in the FWHM beam size of the radiotelescope, or an inclination that appeared on inspection to be less than our limit of $45^{\circ}$.

The region within 3,300 \kms\ is being populated by many other galaxies which are not included in the V3k sample,
yet suitable for the TF relation.  The two samples that will be discussed next make additional contributions to this volume of 639 galaxies 
(PSCz sample) and 440 galaxies (flat galaxies).  Including miscellaneous contributions, we presently are exploring the inner 3,300 \kms\ region with 2,304 TF distance measures.

\subsection{PSCz sample}

We were motivated by two considerations in defining a sample that extends beyond V3K.  First, we now know that our Local Supercluster is just an appendage on a larger supercluster complex that includes the Norma, Centaurus, and Hydra clusters as well as several other clusters such as Abell 3537, 3565, and 3574, all in the so-called `Great Attractor' region \citep{1987ApJ...313L..37D}.  By extending to 6,000 \kms\ we encompass the main part of the overdense region that includes our Galaxy.  It also includes the adjacent Pisces-Perseus filament \citep{1988lsmu.book...31H}.  Our second consideration is the current capability of radio-telescopes.  Satisfactory HI profiles can be obtained for most spiral galaxies within 6,000 \kms\ with integrations of less than an hour with GBT.  The Arecibo Telescope is more sensitive but accesses only a modest fraction of the sky.  The Parkes Telescope is less sensitive and longer integrations are sometimes required but this facility is only required to cover the 15\% of the sky below $-45^{\circ}$. 

In an effort to generate an independent sample that isolates normal spirals and is uniform around the sky, attention was given to the Infrared Astronomical Satellite (IRAS)  Point Source Catalog -- Redshift
(PSCz) 0.6 Jy survey \citep{2000MNRAS.317...55S}.   Extragalactic sources with cool far infrared emission (100 $\mu$m flux greater than 60 $\mu$m flux) are typically normal spirals near morphological type Sc.  Targets can be selected to low Galactic latitudes limited by source crowding.
We selected an all-sky sample of 1,690 targets by the following criteria:

$-$ $V_{cmb} < 6,000$ \kms
 
$-$ IRAS $S_{60} > 0.6$ Jy and $S_{100}/S_{60} > 1$

$-$ inclination from the axial ratio of greater than $45^{\circ}$

$-$ type later than Sa 

$-$ not tidally disturbed

$-$ an HI signal that is not confused

Adequate HI profiles have been acquired for 1204 galaxies in this sample.  The status of observations is shown in Figure~\ref{3samples} where the white histogram describes the initial PSCz sample, the black histogram describes the totality of systems with adequate HI profile measurements, and the grey histogram describes the subset of 470 systems with new HI information from the Cosmic Flows observations.

\subsection{Flat Galaxy sample}

A dominant source of error in the TF methodology arises from inclination measurements.  A characteristic uncertainty is $5^{\circ}$, a contribution of half the error budget through the correction of linewidths toward the face-on limit of $45^{\circ}$.  Worse, there is a tail of large errors with photometrically derived inclinations.  The flat galaxy catalogs compiled from alternatively optical and near infrared images (RFGC: Karachentsev et al. 1999; 2MFGC : Mitronova et al. 2004) provide appealing lists that circumvent this problem.   The extreme axial ratios of the flat galaxy sample assures that candidates are being viewed almost edge-on so uncertainties in the de-projection of circular velocities in a disk are negligible.   With the optically selected component, the axial ratio requirement of a major to minor axial ratio greater than 7 assures that targets have a morphology around class Sc since earlier and later types are intrinsically less thin.  Obscuration issues can be minimized by going to the infrared for the required photometry.   The restriction to extreme edge-on limits the coverage density but results in a coherent and tractable sample for coverage of a large volume.

Our GBT targets from the Revised Flat Galaxy Catalog (RFGC) were restricted to
north of $\delta = -40^{\circ}$ but outside the range
accessed by Arecibo.  Remaining targets within the Arecibo range will either be 
satisfactorily observed with the ALFALFA survey or later with pointed observations 
with the Arecibo Telescope.  Remaining targets at $\delta < -40^{\circ}$ will be
observed with the Parkes Telescope.

In one GBT  semester  we were able to complete most of the missing observations for RFGC galaxies with known radial velocity
less than 6,000 \kms.  Some targets were added to fill in the allocated time windows drawn from the 2MFGC (2MASS-selected Flat galaxy Catalog)  \citep{2004BSAO...57....5M}, or from a list of nearby galaxies for which no modern digital spectrum is available. 
A tiny  fraction of potential targets were a priori discarded, since RFGC galaxies are usually isolated, thus there were a very low 2\% rate of 
confusion in the radiotlescope beam or gross distortion
of the candidate from tidal disruption.  
The non-detection rate was of 14\%: 83 galaxies of 577 were not detected at our flux limit.
More accurately 12\% of the RGFC and 23\% of the 2MFGC galaxies were not detected.
The high detection rate is inherent to the gas-rich nature of the Flat Galaxy sample, while 2MFGC is biased toward 2MASS earlier types galaxies.

The RFGC original sample contains 4,444 galaxies, for which 2,788 have a published redshift (white histogram of the bottom left panel of Figure \ref{3samples}.
Amongst these 2,788 we now have an accurate line width measurement for 1,229 (black histogram).
The Cosmic Flows observations  provided 323 (white histogram) new accurate measurements.

\begin{figure}
\includegraphics[width=8.5cm]{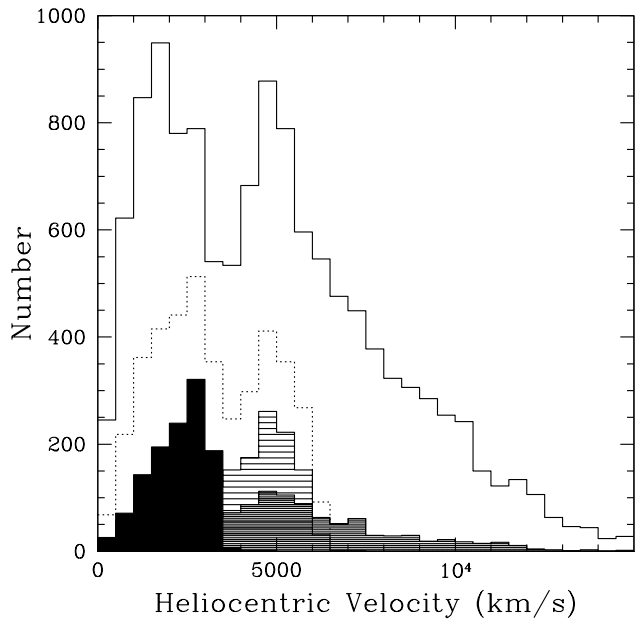}\\
\caption{
Histogram of sample components as a function of velocity.  The V3K sample is illustrated by the filled histogram, the component of the PSCz sample at 3,000-6,000 \kms\ is illustrated by the open hashed histogram, and the fraction of the Flat Galaxy sample with satisfactory HI profile information is illustrated by the denser hashed histogram.  The cumulative samples are outlined by the dotted histogram.  The open solid histogram describes the current accumulation of sources with adequate HI profile information.  
}
\label{hist_nv}
\end{figure}

\begin{figure}
\includegraphics[width=8.2cm]{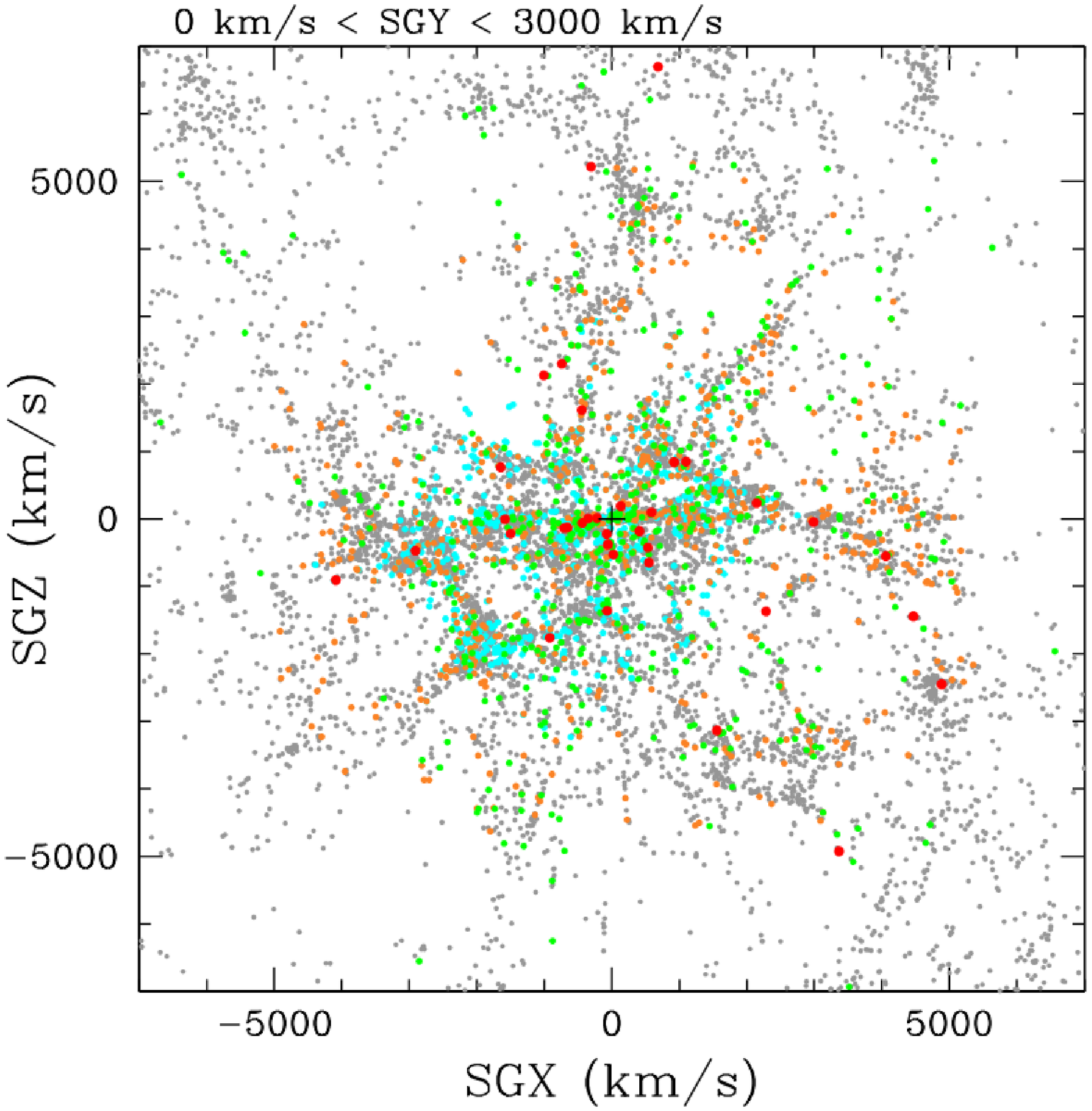}
\includegraphics[width=8.2cm]{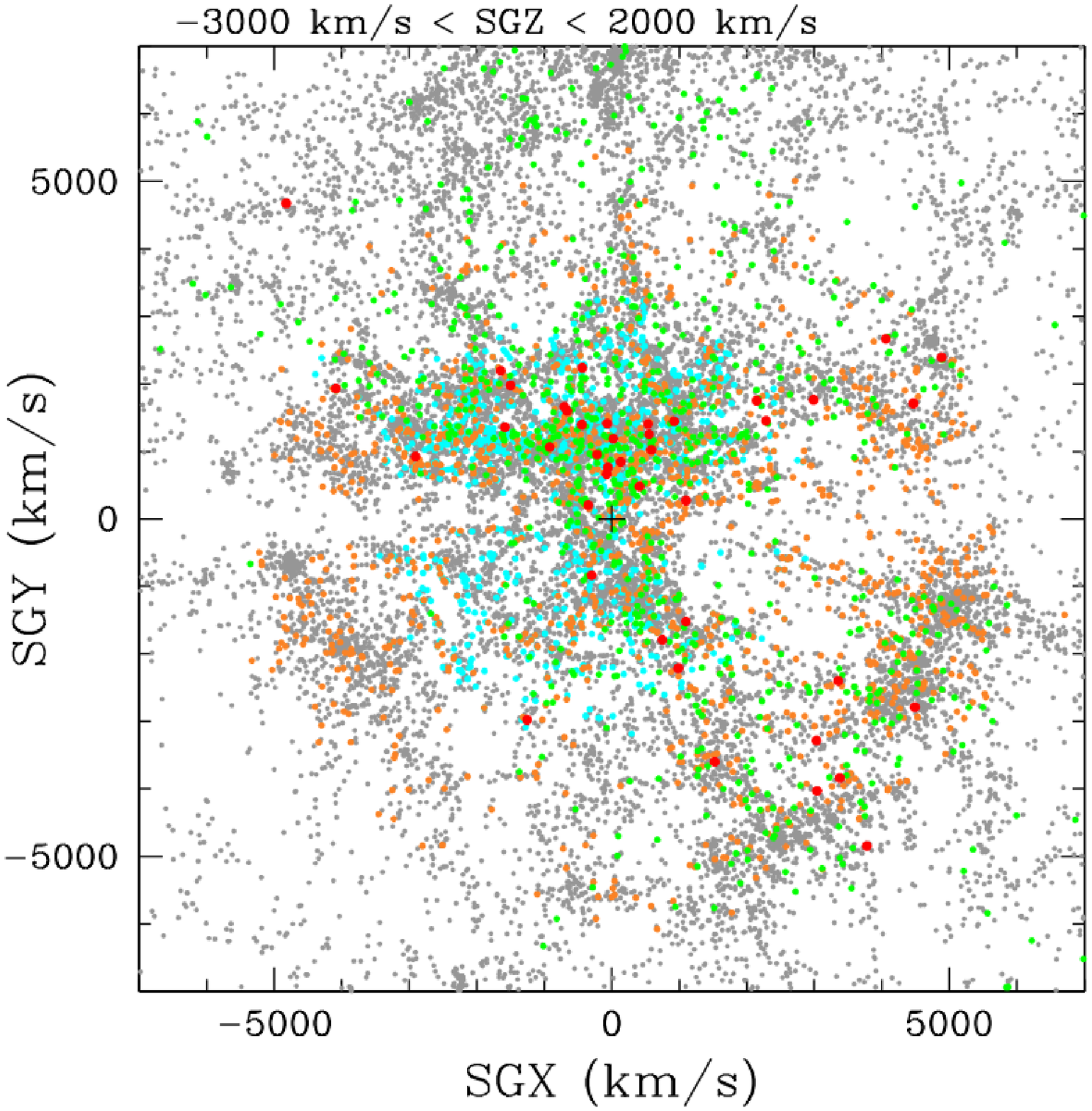}
\caption{
Two slices in supergalactic coordinates that illustrate the coverage of the various samples. {\it Top:} A 3,000 \kms\ thick slice in SGY that shows the main structure of the Norma-Hydra-Centaurus-Virgo supercluster complex.  {\it Bottom:} A 5,000 \kms\ thick slice in SGZ that includes the main features of our supercluster complex and includes the Pisces-Perseus structure at negative SGY and positive SGX. 
All known galaxies from a redshift catalog are plotted in grey.  The V3K sample is over-plotted in cyan; the PSCz sample is over-plotted in orange; the Flat Galaxy sample is over-plotted in green; and the SNIa sample is over-plotted in red in larger symbols.  
}
\label{xz_xy}
\end{figure}

Two figures summarize the properties of the various samples.  Figure~\ref{hist_nv} gives combined histograms of the samples.  There are large numbers of sources within the formal samples and large numbers of additional sources with good HI profile measurements within 3,000 \kms.  At $3,000-6,000$ \kms, there are comparable numbers of sources to the inner 3,000 \kms\ though spread through a larger volume.  Beyond, there is a tail of rapidly decreasing coverage that extends to $\sim 12,000$ \kms.  The distribution of many of these sources can be seen in the two slices of a three-dimensional cube in Figure~\ref{xz_xy}.  It is seen that the nested samples provide good coverage of our immediate over-dense region but falls off rapidly beyond the edges of the structure that we live in.

\section{Observations with Green Bank Telescope}

To strive for completion of the V3K, PSCz, and Flat Galaxy samples, we have been observing with the 100m Green Bank Telescope at declinations above $\delta = -45\degr$, the southern limit with GBT, but excluding the Arecibo range $0 < \delta < +38$  where data is currently acquired within the ALFALFA project.  Our present strategy is to await the results of the ALFALFA
multibeam survey currently in progress.  Currently ALFALFA data releases $1-3$ have been processed through our pipeline.
 If profiles are inadequate we will
request deeper observations of those targets in the Arecibo range of declinations.
Access to the remaining sky, at $\delta < -45\degr$, requires observations with the Parkes Telescope in Australia.  Sources are 
not observed if digital spectra of good quality exist. 

The single-beam Robert C. Byrd Green Bank Telescope (GBT) observations released in this paper 
are identified as ctf2009 and ctm2010 in our "All Digital HI catalog".
Observations were carried out spanning  from February 2007 to June 2010 with project identifiers: 07A039 (55 hrs), 07C067 (218 hrs), 08A072 (47 hrs), 08B041 (50 hrs), 08C010 (340 hrs), and 10A059 (370 hrs). 
The total of observing time allocated to the project at GBT was 1080 hrs.
GBT has implemented a "Legacy ID" numbering for the public access of data, our project data can thus been recovered under
these Legacy ID: GF13, GC47, GC60, GC67, GC69, GC102.

For the observations we use the single beam L band (1 to 2 GHz) receiver
and the spectral line spectrometer as the backend detector.
The final spectrum is stored with 1.6 \kms\ resolution. It was usually binned at least once to 3.2 \kms\ resolution for the HI  linewidth measurement.

\begin{figure}
\includegraphics[width=7.5cm]{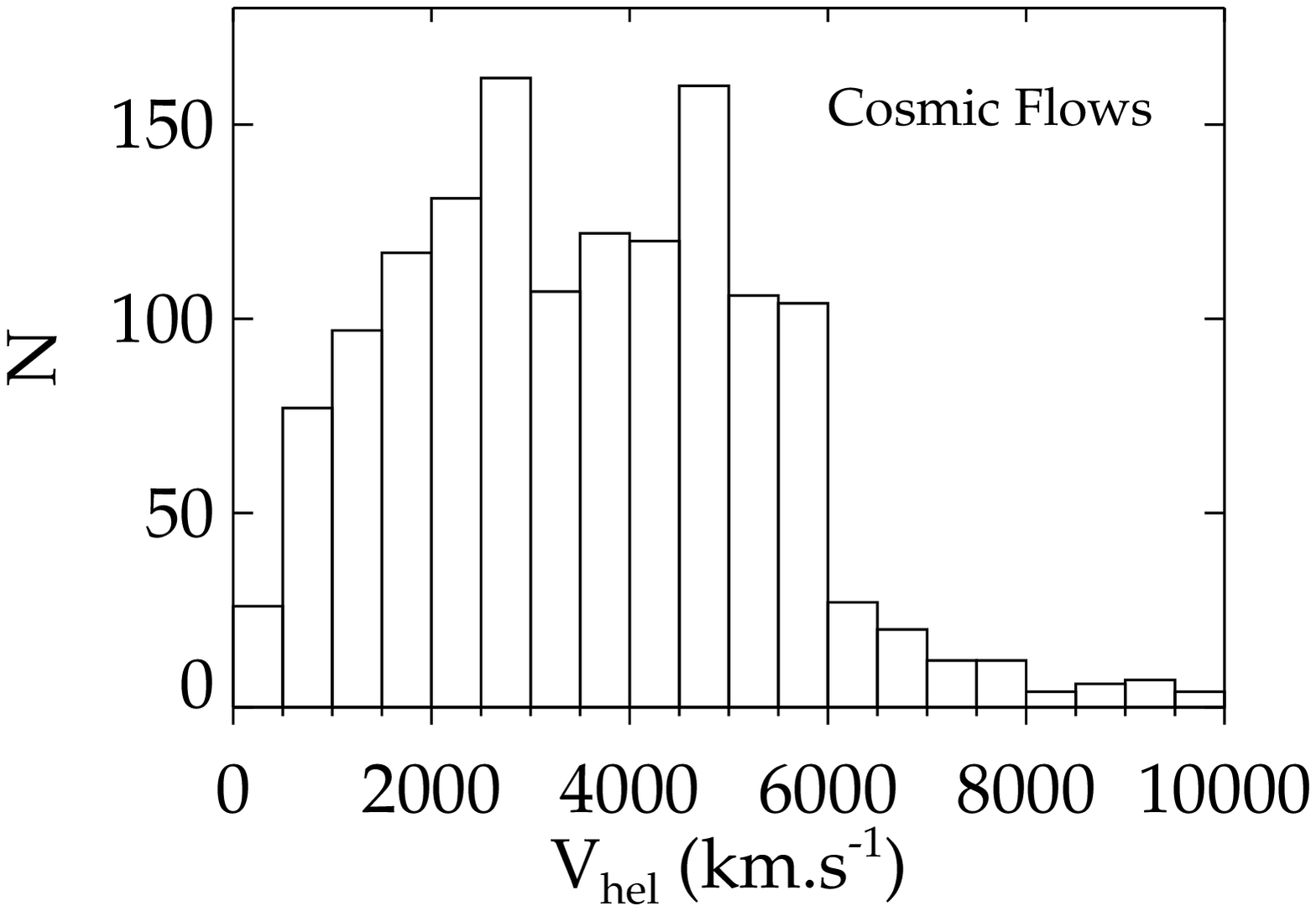}\\
\includegraphics[width=7.5cm]{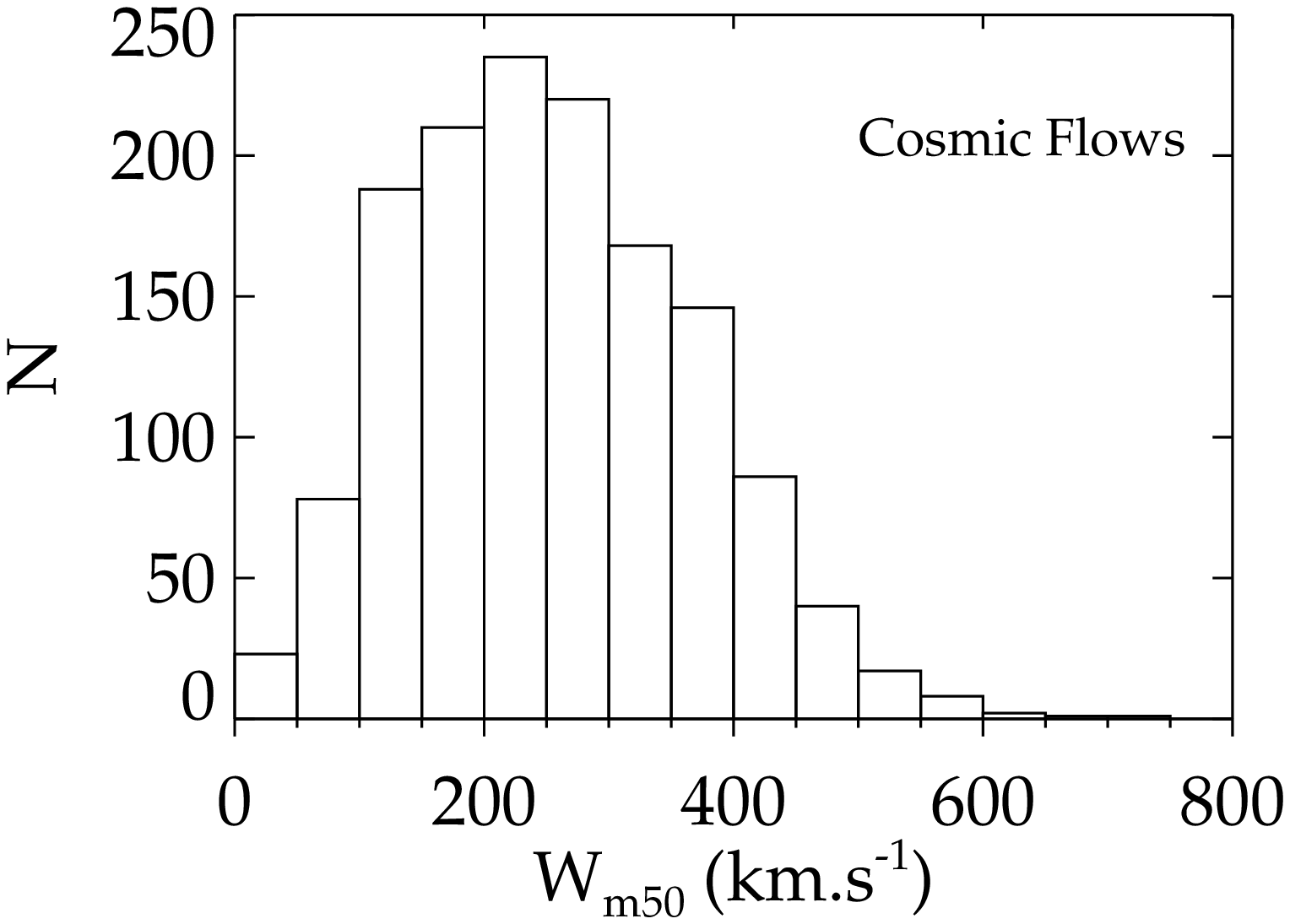}\\
\includegraphics[width=7.5cm]{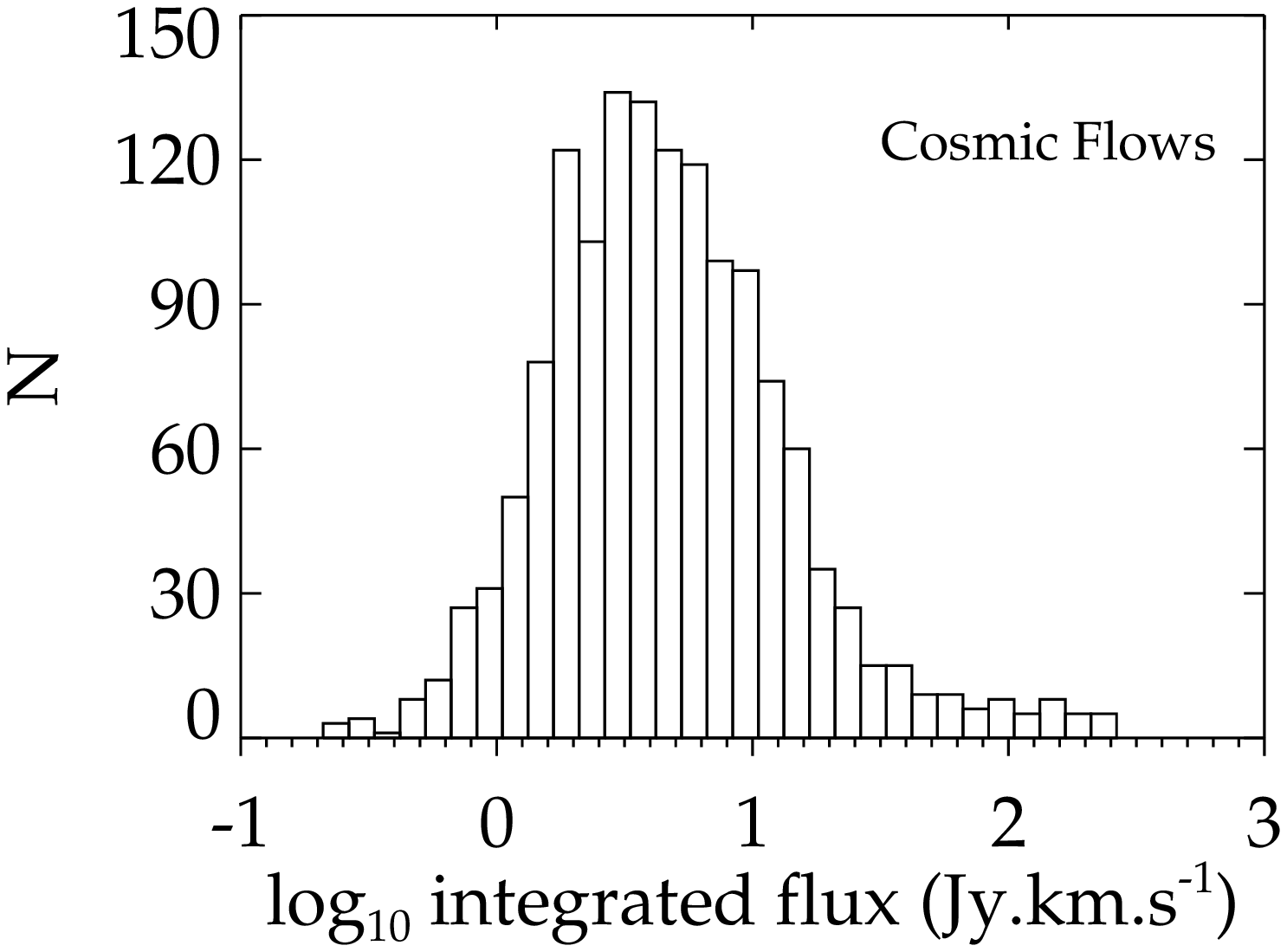}\\
\includegraphics[width=7.5cm]{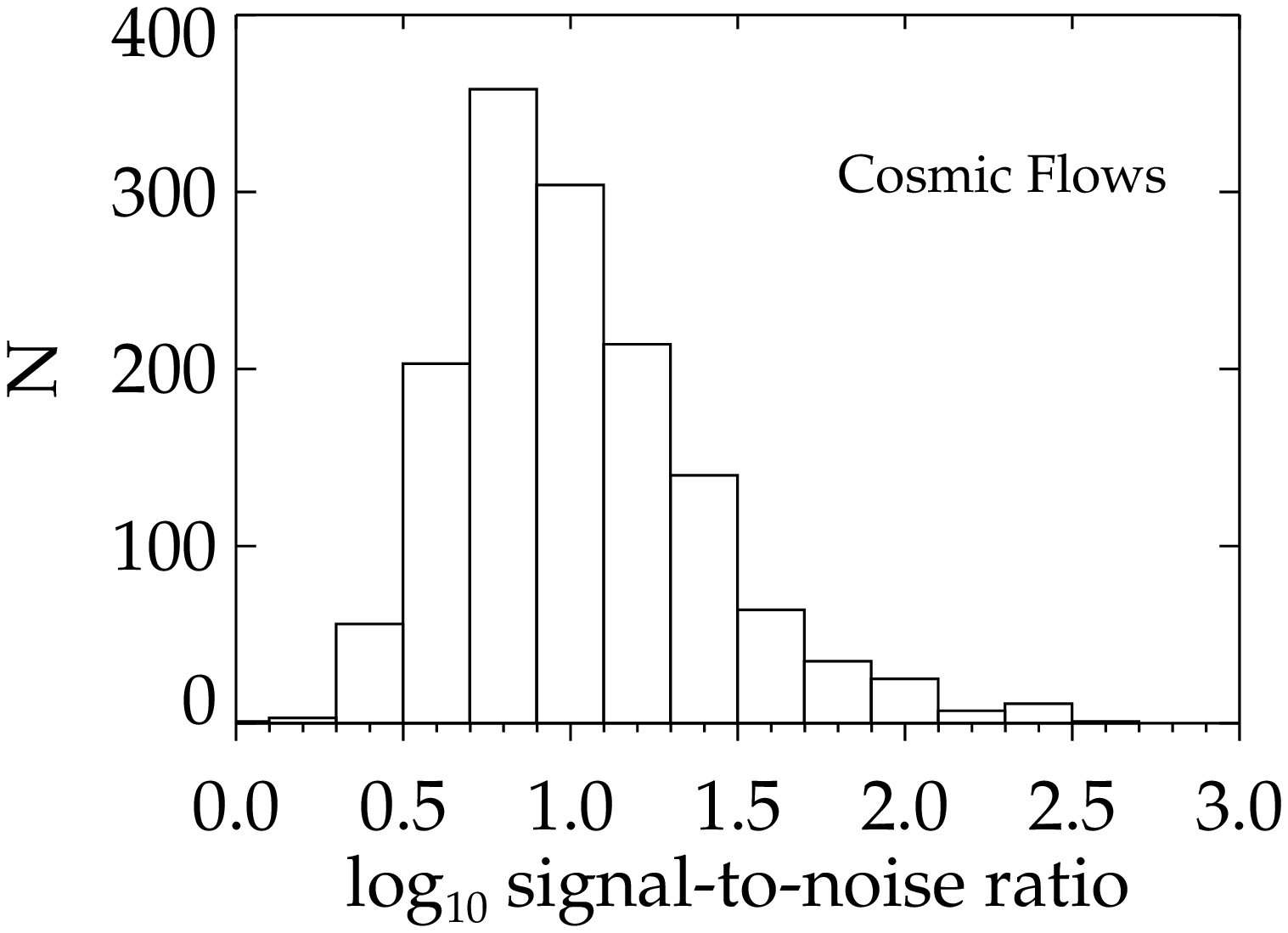}\\
\caption{
Histograms of the galaxy properties of the  1,423 suitable profiles obtained in the Cosmic Flows Program in HI at GBT and Parkes:
(a) heliocentric recession velocity in \kms; (b) HI linewidth at half power (W$_m50$); (c) logarithm integrated flux ; (d) logarithm of the signal-to-noise ratio.
}
\label{hist_courtois}
\end{figure}

\section{Observations with the Parkes Telescope}

Southern targets $\delta <$ -45 deg for the two samples V3K and PSCz are currently being observed with 
the Parkes Telescope.  

The Parkes 64m radiotelescope observations released in this paper  were carried out in February 2009, 1-10: 
P660: 60 hrs on the sky. 
We were able to observe 58 galaxies with known radial velocity
less than 4,000 \kms.  We obtained 33 spectra leading to accurate linewidths measurements and 24 are inadequate for distance measurements.
A typical exposure time for a target at 3,000 \kms\ was 1.5 hour or more. The non-detection rate was 16\%: 9 galaxies of 58 were not detected.

Figure~\ref{hist_courtois} illustrates the general properties of the HI profiles that were adequately detected in the course of the Cosmic Flows Large Program with GBT and the supplement with Parkes Telescope.

\section{ Analysis of HI Line Profiles}

With the award of our project as an NRAO GBT Large Program, we reconsidered the traditional ways
of measuring the HI linewidth. We were looking for a robust method that could be applied to both low and high signal-to-noise ratio spectra.
We also wanted to be able to remeasure a large quantity of archive HI data from various telescopes, so the method would be as automated as possible.

Historically, Tully and Fisher measured linewidths $W_{20}$ at 20\% of peak intensity \citep{1977A&A....54..661T}.
However the 20\% level is frequently close to the noise level leading to spurious measurements.
Also the 20\% level is difficult to secure when profiles are strongly asymmetric.  
We now prefer to measure the width 
enclosing 50\% of the cumulative HI line flux, $W_{m50}$.   Specifically, $W_{m50}$ is the line width at a flux level that is 50\% of the mean flux averaged in channels within the wavelength range enclosing 90\% of the total integrated flux.  The exclusion of 5\% of the flux at the high and low wavelength edges minimizes problems in the profile wings.  The $W_{m50}$ line widths are measured at slightly higher flux levels than $W_{20}$.  The random scatter between these two measures is the lowest found among alternatives that were examined.  

The parameter $W_{m50}$ is an empirical measure of the width of an HI profile.  We correct for redshift and instrumental broadening with the formula
\begin{equation}
W_{m50}^c = {{W_{m50}} \over {1+z}} - 2 \Delta v \lambda
\label{Wc}
\end{equation}
where $z$ is the redshift, $\Delta v$ is the smoothed spectral resolution, and $\lambda=0.25$ is an empirically determined constant.
The observed line width can also be adjusted by separating out the broadening from turbulent motions and offsetting to produce an approximation to $2 V_{max}$ where $V_{max}$ characterizes the rotation rate over the main body of a galaxy.  We have defined the parameter $W^i_{mx}$ where
\begin{eqnarray}
\nonumber
W_{mx}^2  = W_{m50}^2 + W_{t,m50}^2 [1 - 2  e^{-(W_{m50}/W_{c,m50})^2}] \\
 -  2 W_{m50} W_{t,m50} [ 1 - e^{-(W_{m50}/W_{c,m50})^2}]  
\label{WR}
\end{eqnarray}
with $W_{c,m50} = 100$~\kms\ and $W_{t,m50} = 9$~\kms.  Then $W^i_{mx} = W_{mx}/{\rm sin}(i)$ where $i$ is the galaxy inclination from face on.
Details regarding the $W_{m50}$ and $W^i_{mx}$ line width parameters and comparisons with alternatives are discussed by  \citet{2009AJ....138.1938C}.

All modern spectra are available in a digital form.  Consequently it is straight
forward for us to pass archival data through the same analysis procedure as we
apply to our newly observed material.  It, of course, means that we do not need to observe
a galaxy if we can access adequate observations from any archive.  The primary
sources of data in our current catalog originate, by decreasing number, from Arecibo (7,898), Nan\c{c}ay (3439),
GBT (1,444), Parkes (1052),  the old NRAO  $300^{\prime}$ (1,059), the $140^{\prime}$ (696)
and Effelsberg (235).

\begin{figure*}
\centering
\begin{tabular}{cc}
\epsfig{file=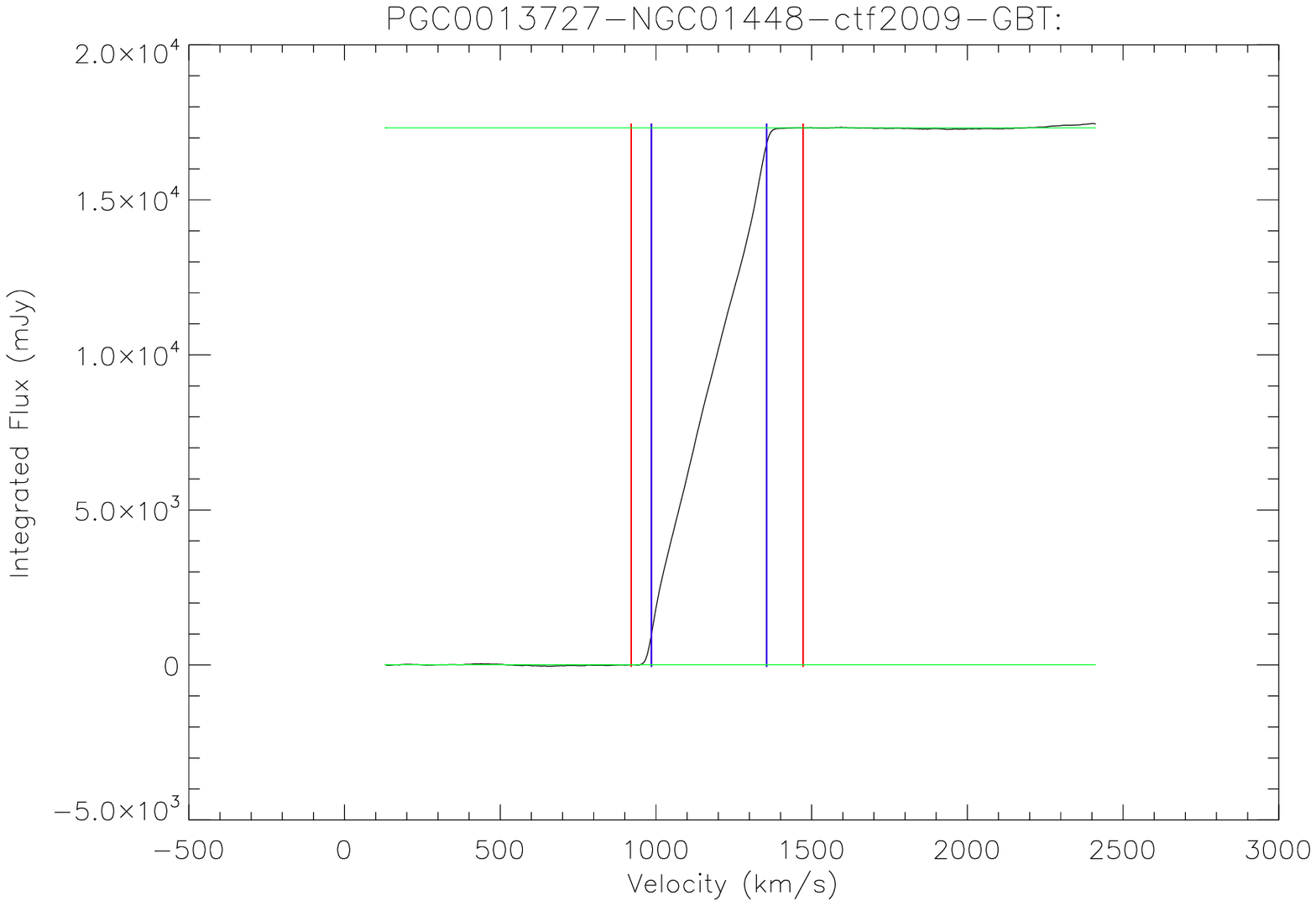,width=0.49\linewidth,clip=} & 
\epsfig{file=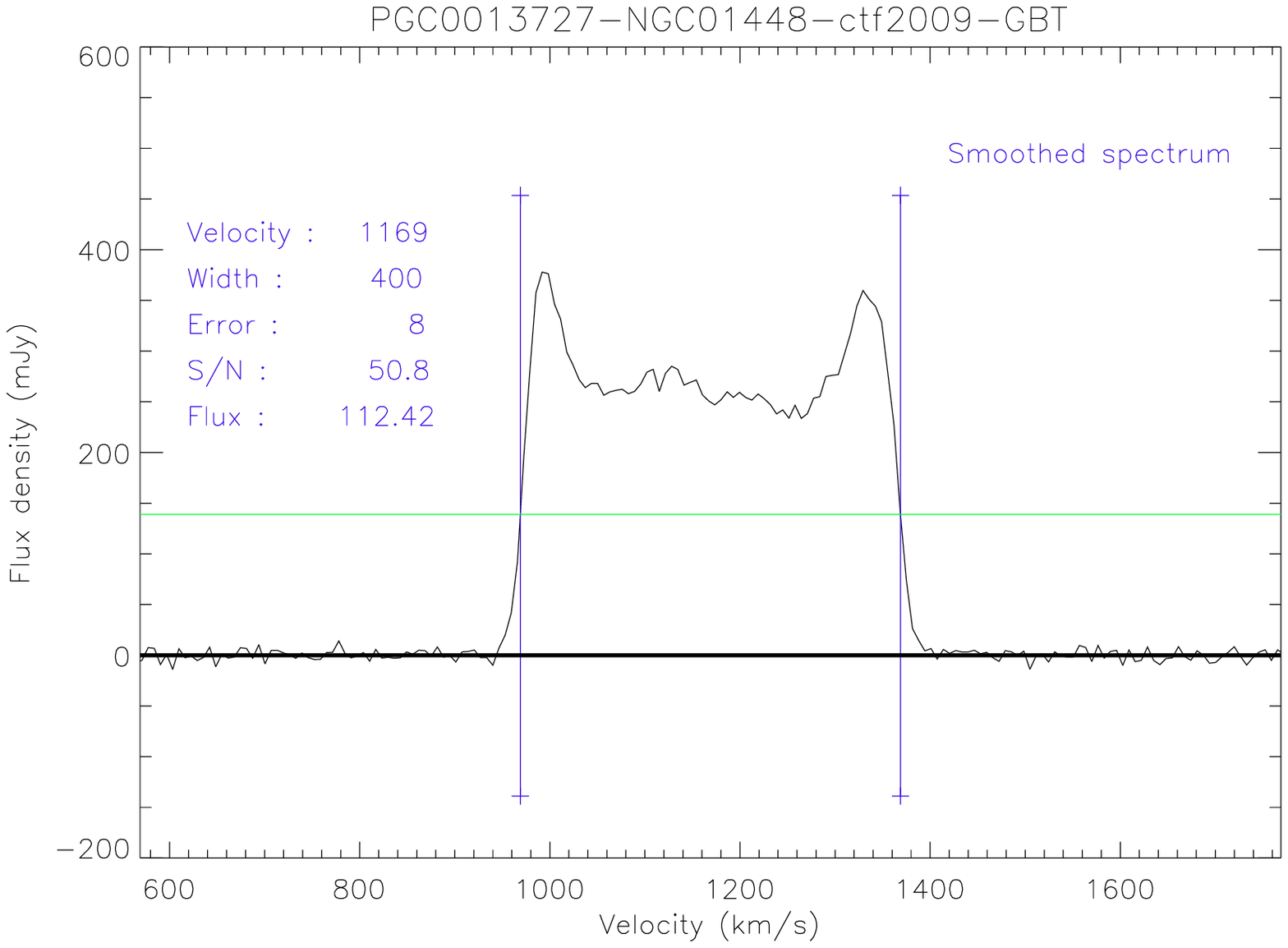,width=0.49\linewidth,clip=} \\
\epsfig{file=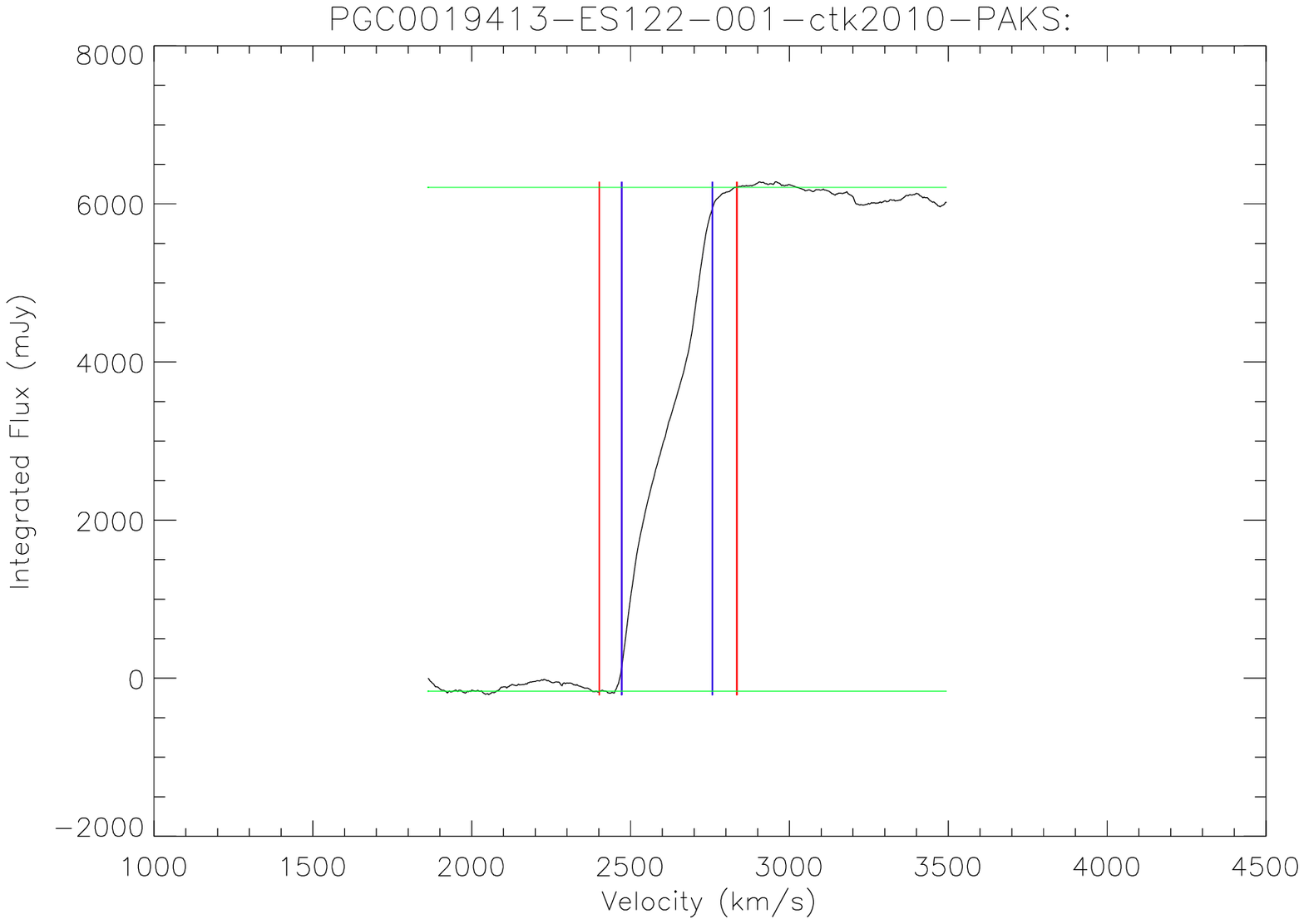,width=0.49\linewidth,clip=} &
\epsfig{file=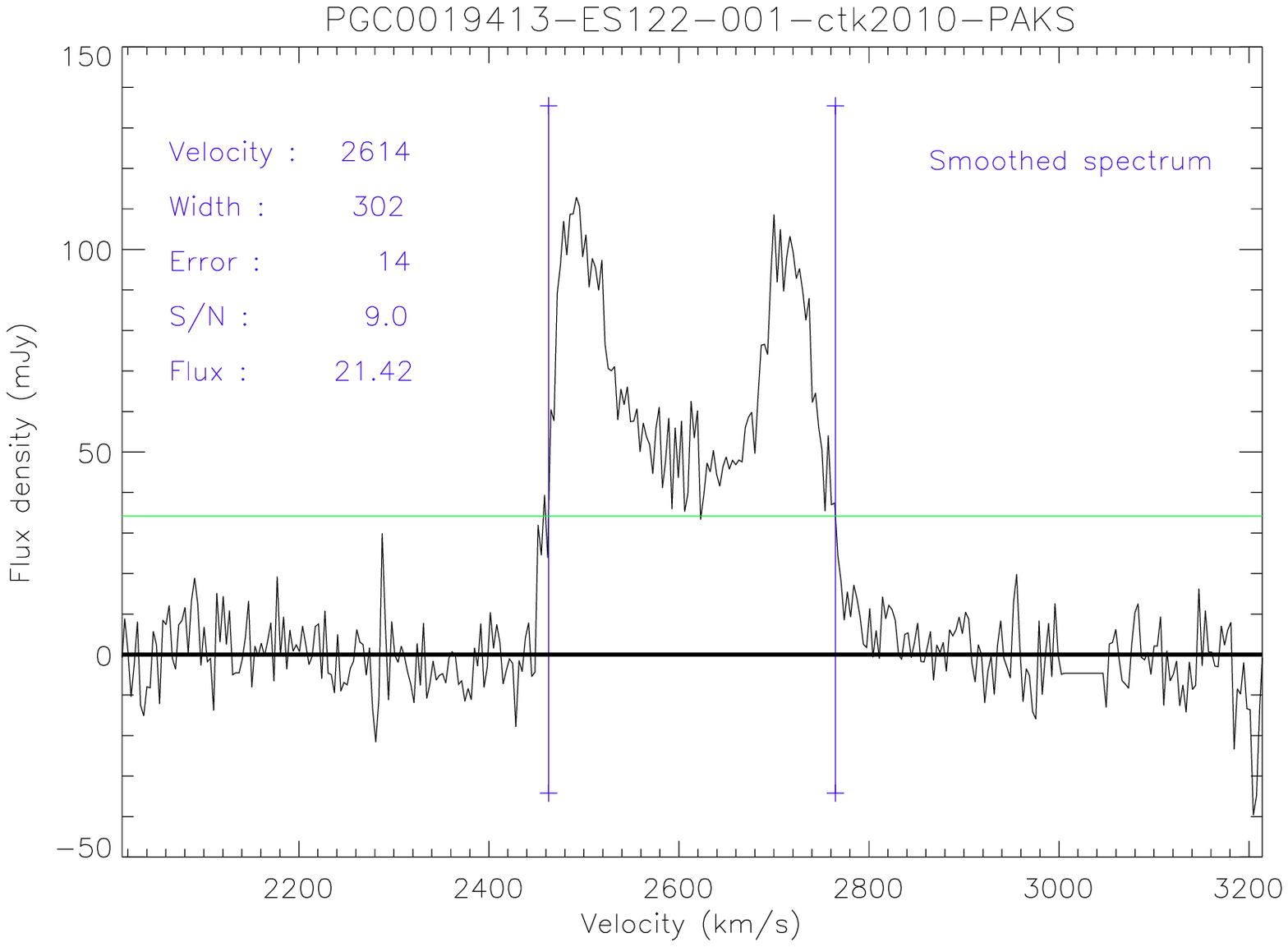,width=0.49\linewidth,clip=}\\
\epsfig{file=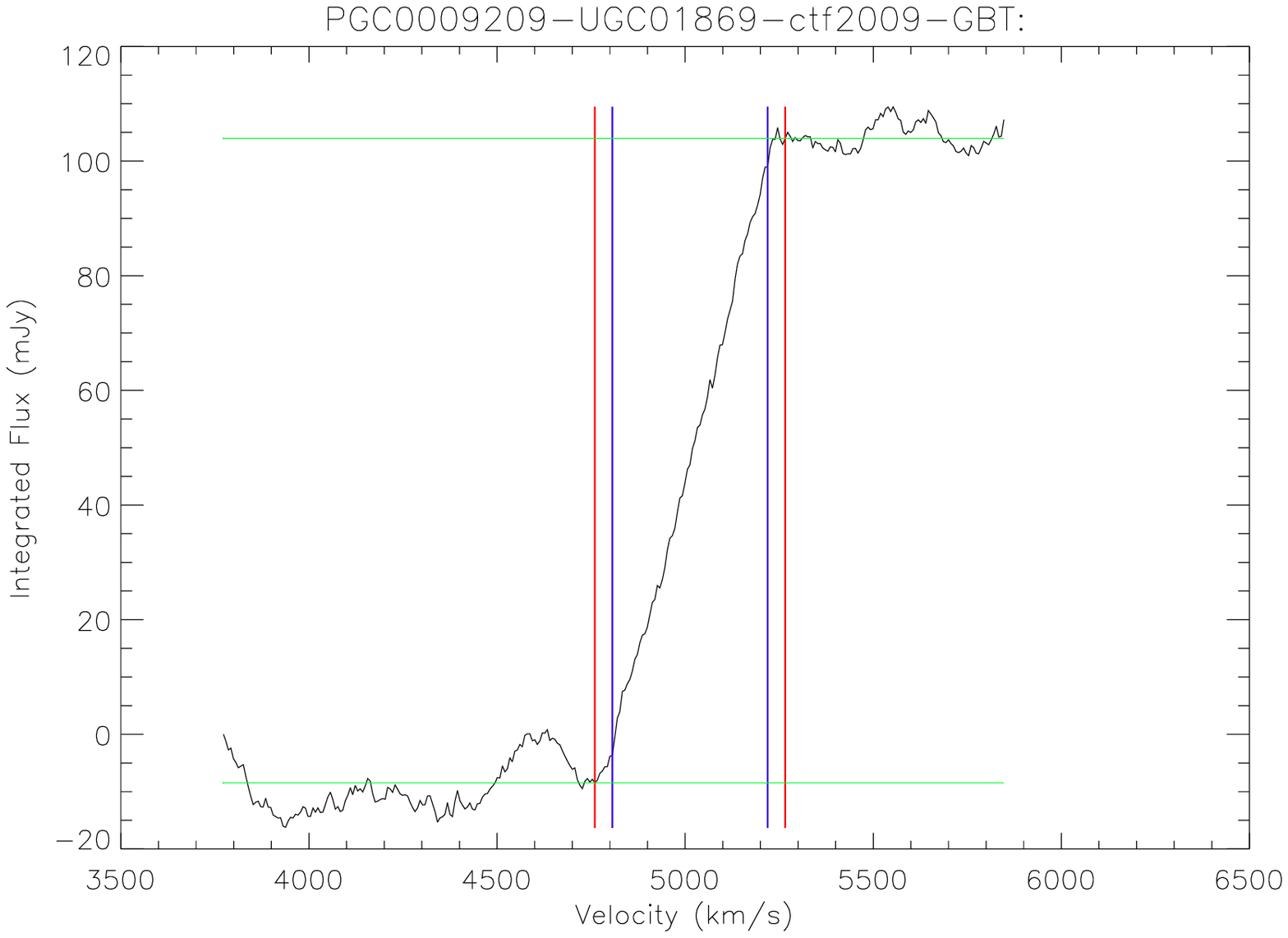,width=0.49\linewidth,clip=} &
\epsfig{file=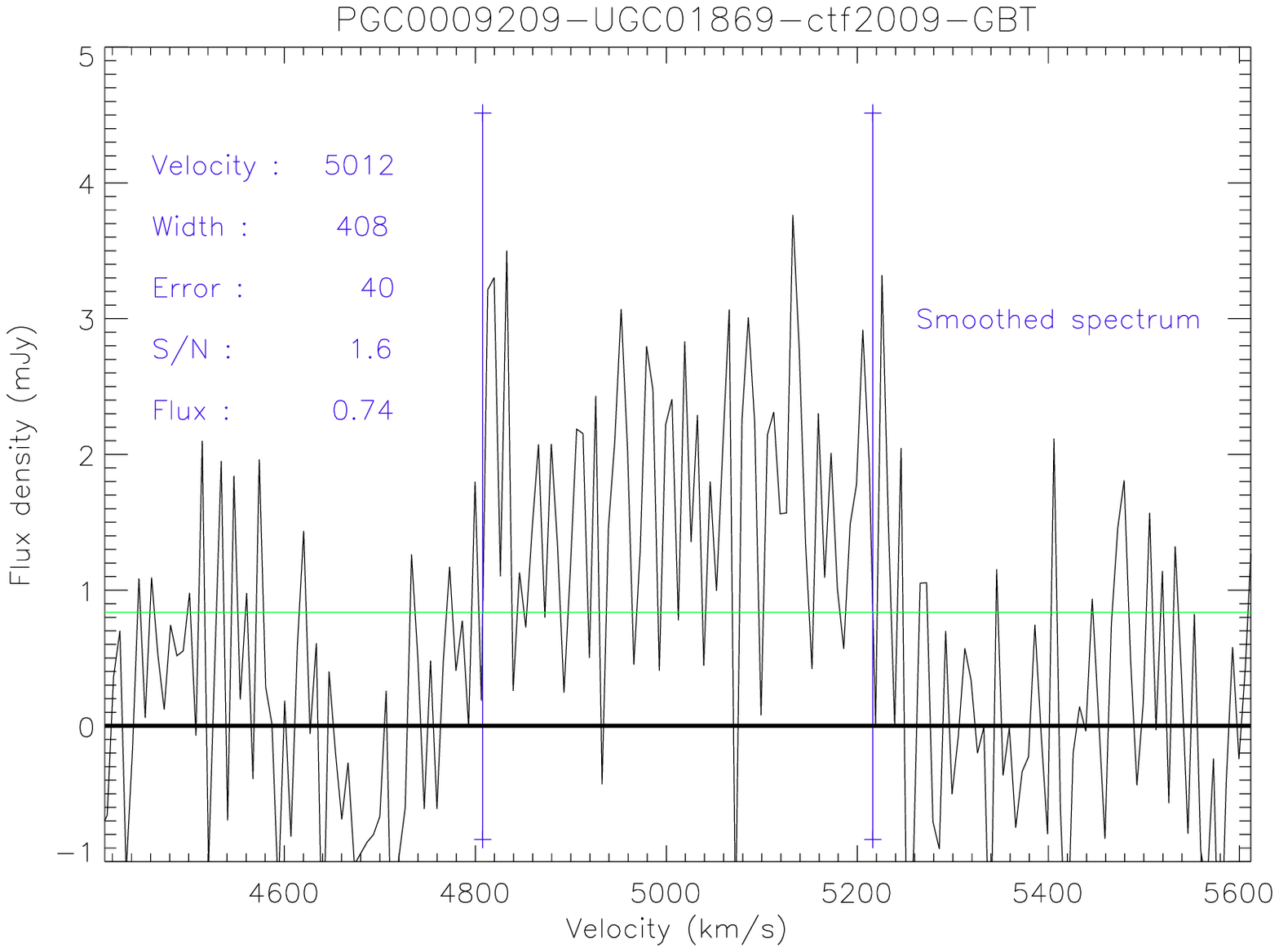,width=0.49\linewidth,clip=} \\
\end{tabular}
\caption{
Example of two adequate profiles (top and middle) and one inadequate (bottom). Profiles are suitable for distance measurement using the Tully-Fisher Relation when an error on W$_{m50}$ was set to 20 \kms or less.
The corresponding flux integration curves are displayed on the left with the 5\% and 95\% window (blue vertical lines) for the W$_{m50}$ measurement. These profiles are showing an icreasing error from top to bottom.}
\label{adequate}
\end{figure*}

The assignment of errors for $W_{m50}$ or quality flags is a particularly challenging problem.  Formally, we link errors to the profile signal to {\it rms} noise as described by \citet{2009AJ....138.1938C}.  From comparisons between alternative observations our cited errors appear to be conservative; about 50\% larger than a $1 \sigma$ value.  Also on the side of being conservative, we assume that the line width uncertainty is 8~\kms\ in the best of cases.  Estimates are difficult to quantify precisely because the error on $W_{m50}$ is  linked to three characteristics:  signal-to-noise ratio, observational flux limit, and shape of the HI line with flux in the wings that may or may not be real.  Our interest is to use the profile widths as a parameter in the measurement of distances.  Even if our errors do not have a precise absolute sense they do have a relative sense.  Within our system, we consider that there is a threshold of acceptability: a profile may be of sufficient quality to be used in the determination of a distance, or it may not be.  We link our error estimate to this threshold.  Specifically, an adequate profile is assigned an error of less than or equal to 20~\kms.  Inadequate profiles are identified by errors greater than 20~\kms\ (see examples in Figure~\ref{adequate}).  Confused profiles are identified by the error flag of 100~\kms. Non-detections are identified by the error flag of 500~\kms.
Since for distance measurement purposes we will not use any profile with an error larger than 20 \kms, 
measurements with such large errors are not taken into account in the averaged  $W_{mx}$\_av column of our `All Digital HI Catalog'. Thus if a galaxy has 
only one measurement and it is considered inadequate then there will be a value given for $W_{m50}$ among the parameters measured from the specific source profile but no value will be registered for $W_{mx}$\_av, see for example line 2 of Table \ref{HIobs}.

In our compilation of digital spectra, some galaxies have been observed by 2 or 3 different telescopes.  Inter comparisons between different HI observations of the same galaxies suggest that the characteristic accuracy of an individual acceptable profile width is 7~\kms.

Our HI observations at Green Bank  and Parkes 64m telescopes are made available in an electronic form with this article. A few lines are given in Table \ref{HIobs} as example. The Columns are respectively: PGC name, source of observations, telescope, heliocentric velocity, W$_{m50}$, error on W$_{m50}$, signal-to-noise ratio, integrated flux in the line.
\begin{table*}
\caption{1,822 Cosmic Flows Green Bank and Parkes HI observations}
\label{HIobs}
\begin{tabular}{lllccccc}
\hline
PGC  &source & telescope & V$_{hel}$ & W$_{m50}$ & error on W$_{m50}$ & signal-to-noise& integrated flux\\
name      &  code    &     & km/s  & km/s & km/s & ratio & Jy.km/s \\
\hline
PGC0000094   & ctm2010 &   GBT &  4098  &188   &14 &   8.9 &   3.6\\
PGC0000207   & ctf2009   & GBT  & 6597   &    & 25 &   2.6   & 1.9\\
PGC0000218   & ctf2009    &AOG  & 1051 & 453   & 8  & 26.2 &  11.7\\
PGC0000279  &  ctm2010   & GBT  & 2311&  411  & 11  & 12.4  & 14.0\\
PGC0003743  &  ctk2010  & PAKS &  2310 & 361 &  17  &  5.0 &  13.5\\
\hline
\hline
\end{tabular}
\end{table*}

\section{Linking Analog and Digital Line width Measurements}

Even though there are now digital spectra for most galaxies the collection remains incomplete.  Indeed, it will be difficult to achieve completion.  Some very nearby galaxies are much larger than the FWHM beam sizes of modern radio telescopes.  They could be mapped.  Better, integrated profiles can be reconstructed from observations with the Westerbork, Very Large Array, or Australia Telescope interferometers.  Alternatively, profiles obtained in earlier days with smaller telescopes like the NRAO $140^{\prime}$ or the Dwingeloo facilities might be the best available.  

A considerable effort was made in earlier years to accumulate a consistent compilation of analog line widths.  The contributing observations were made with many telescopes and reported by many sources but all  $W_{20}$ measures in our compilation were coherently measured by only three people: J.R. Fisher, R.B. Tully, or C. Hall.  

\begin{figure}
\includegraphics[width=8.5cm]{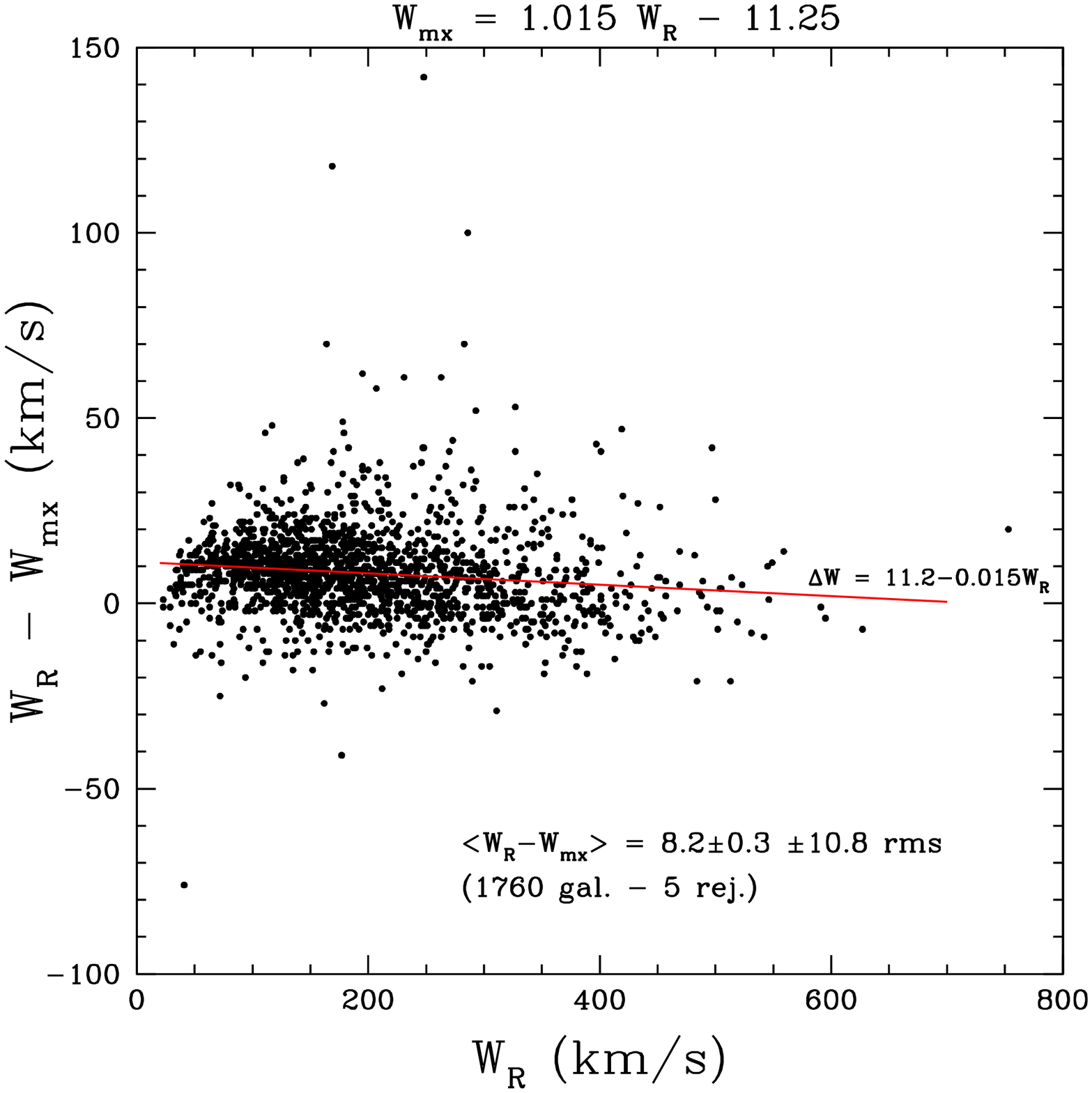}
\caption{
Comparison of the line width measures designed to approximate $2 V_{max}$ with alternatively the old analog measurements and the new digital measurements of line width, respectively $W_R$ derived from $W_{20}$ and $W_{mx}$ derived from $W_{m50}$.   Globally there is an offset in the difference of the two parameters of 8~\kms\ with a dispersion of 11~\kms.  In detail there is a small gradient in the difference described by the solid line.
}
\label{W_all_pre}
\end{figure}

A comparison was made between analog $W_{20}$ measures and our new digital $W_{m50}$ measures by \citet{2009AJ....138.1938C}.  In applications related to distance determinations we will be more interested in the line width parameters related to the physical property $V_{max}$, the characteristic rotation velocity across the disk of a galaxy.  To this end, we now compare the parameters that approximate $2 V_{max} {\rm sin} i$ for the separate analog and digital samples.  This comparison is shown with Figure~\ref{W_all_pre}.   The analog parameter $W_R$ is calculated here using a constant value $W_{t,20} = 22$~\kms\ in the equivalent to Eq.~\ref{WR} for the case of the analog transformation using $W_{20}$.   This value for the thermal broadening constant is different from the value $W_{t,20}^{old} = 38$~\kms\ originally advocated \citep{1985ApJS...58...67T} for the transformation to $W_R$.  For a discussion of this change see \citet{2009AJ....138.1938C}.  Based on 1755 galaxies with good analog and digital profiles, we find a relation that allows us to transform from analog to digital parameters:
\begin{eqnarray}
W_{mx} = 1.015 W_R -11.25
\label{WRtoWmx}
\end{eqnarray}
The r.m.s. scatter is 10~\kms\ suggesting uncertainties of roughly 7~\kms\ in each of the measured parameters.  

It is seen that there is a slight slope to the difference between the alternative line width measures $W_R$ and $W_{mx}$ which in principal are both approximating $2 V_{max} {\rm sin} i$.  The two measures are essentially equal at large line widths but there is a mean difference that grows to 10~\kms\ for the smallest line widths.  Because of this systematic between $W_R$ and $W_{mx}$ it is essential that the relation between galaxy luminosity and line width (the TF relation) be calibrated for the specific line width measure that is used.  Whatever the line width measure and to whatever degree it may approximate a physical parameter, it remains subject to observational vagaries. 

\section{HI Catalog in Extragalactic Distance Database(EDD)}

Our new observations from the Cosmic Flows program and archival information are combined.    Tabular information and line profile plots are provided for 13,941 galaxies in the All Digital HI Catalog (ADHI) in EDD.  Intensity-velocity ascii tables are provided in the cases of material from the Cosmic Flows observational program.  Currently the catalog contains 16,004 HI profiles, including 1,859 new profiles added through this program.  There are 11,074 profiles that are deemed acceptable; ie, with uncertainties $\le 20$~\kms.   For 1339 galaxies there are at least 2 acceptable profiles and in 82 cases there are three acceptable profiles.  Additions are continually being made to the catalog.

\begin{figure*}
\centering
\begin{tabular}{cc}
\includegraphics[width=8.3cm]{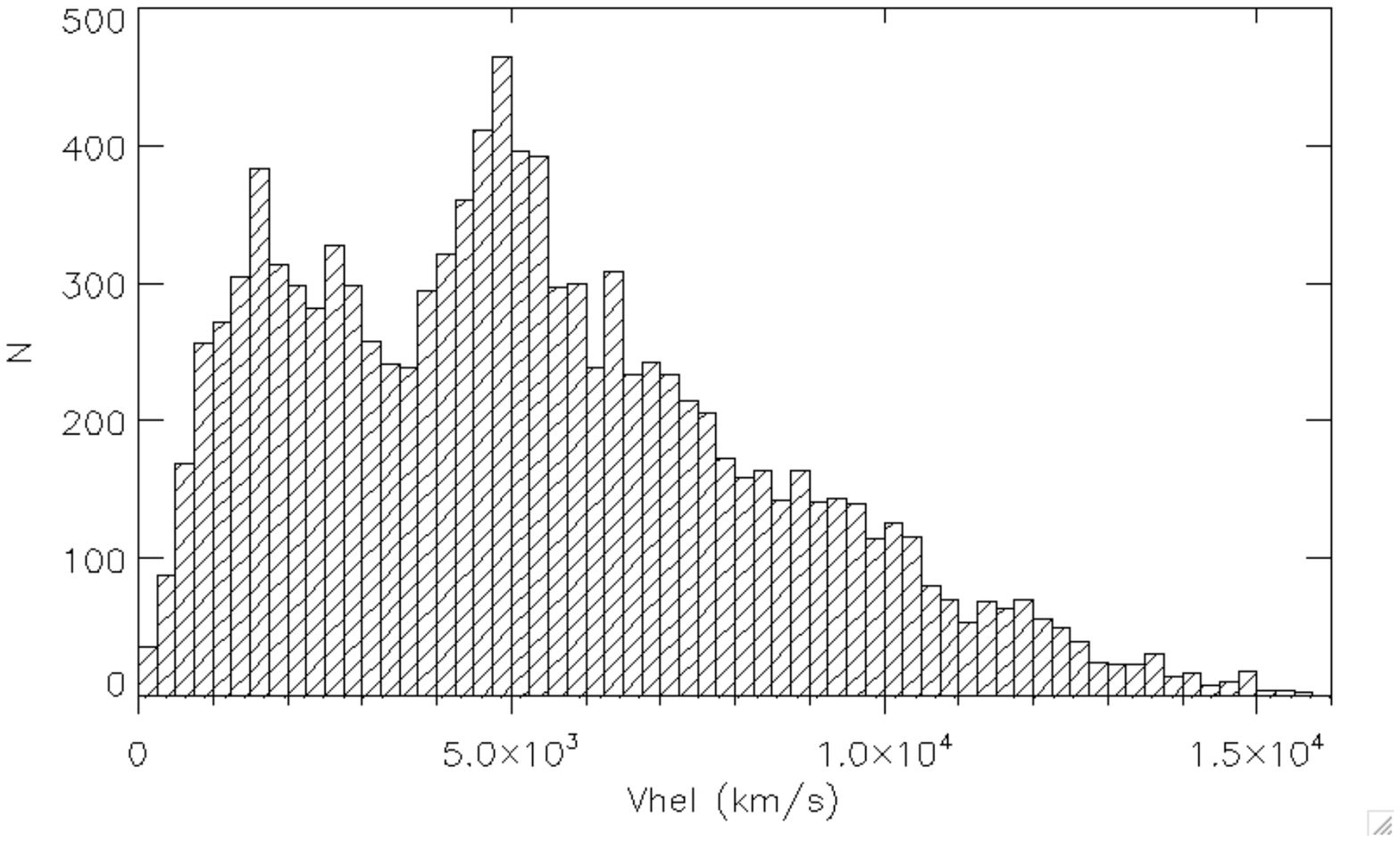}&
\includegraphics[width=8.3cm]{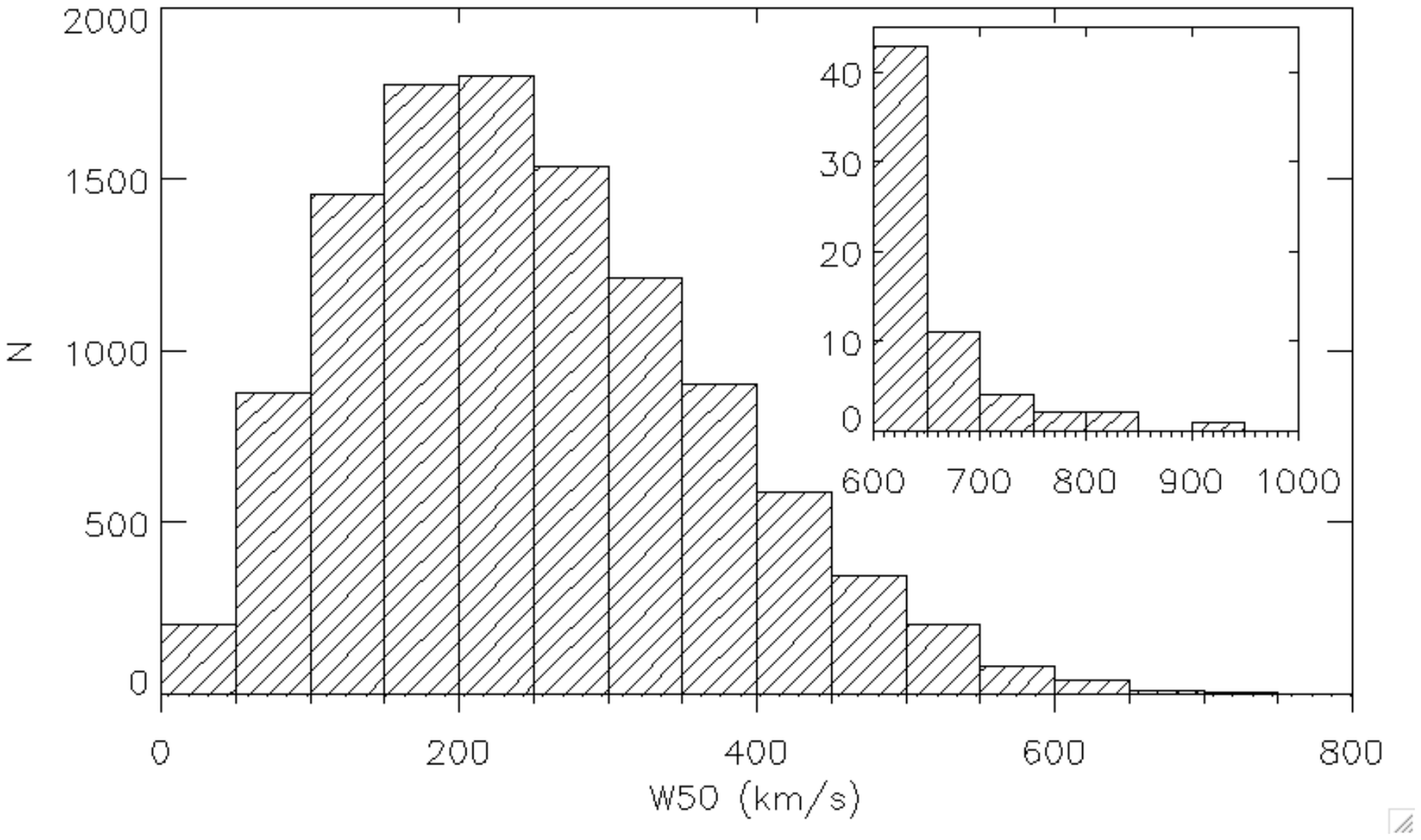}\\
\includegraphics[width=8.3cm]{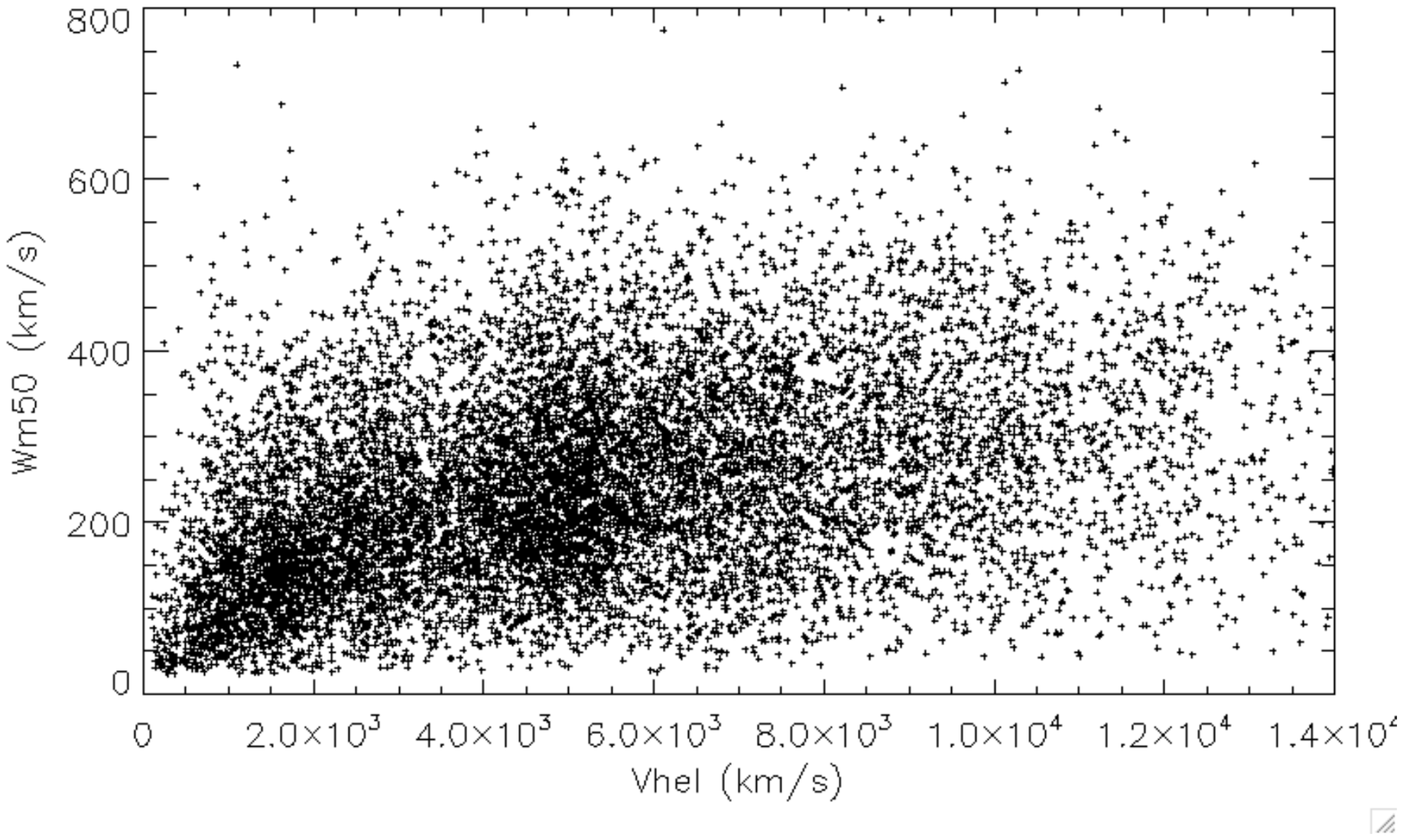}&
\includegraphics[width=8.3cm]{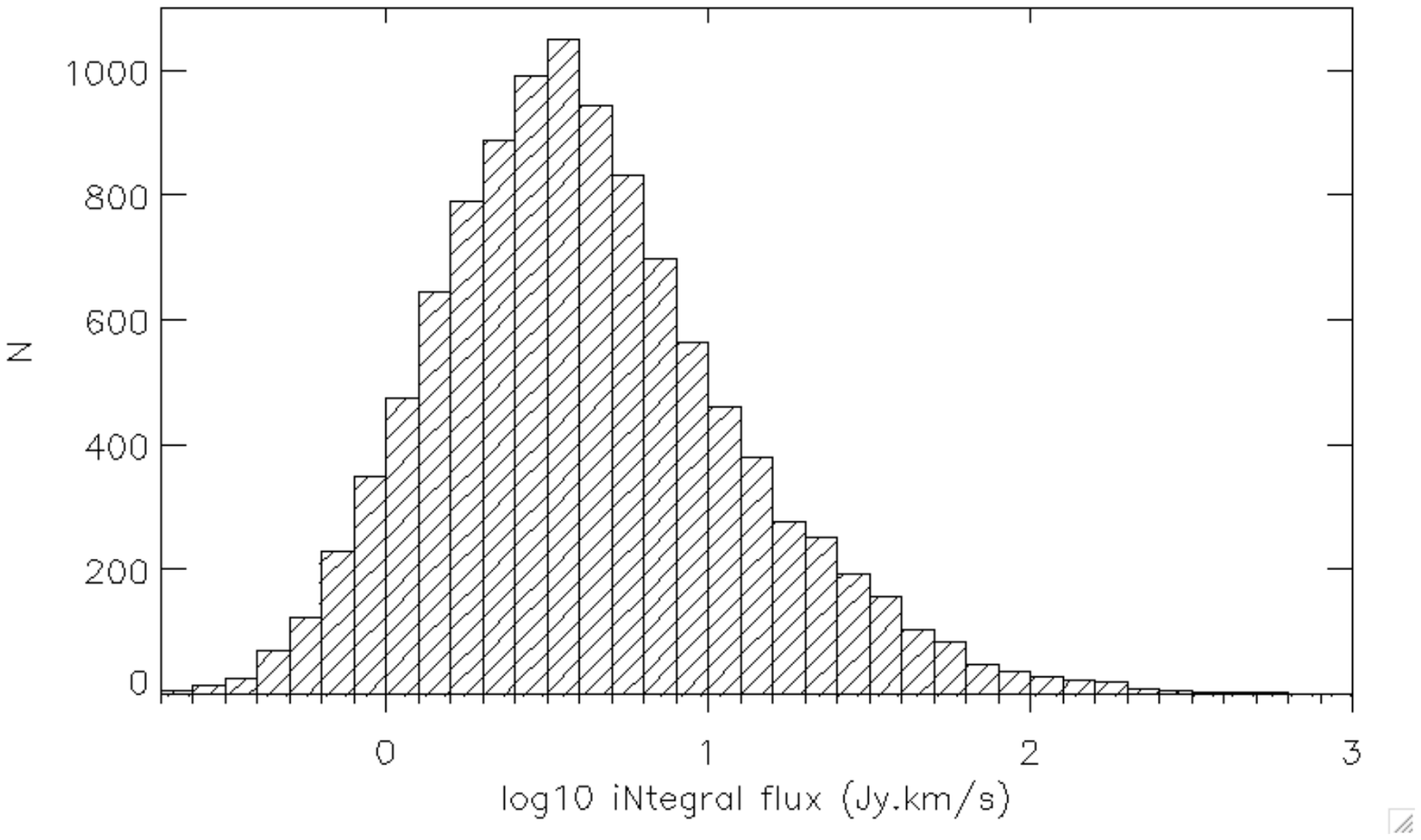}\\
\end{tabular}
\caption{
Histograms of the galaxy properties of the 11,051 galaxies with adequate HI profile for distance purpose which have been consistently analyzed by the Cosmic Flows project:  
distribution in  heliocentric recession velocity in \kms; distribution in HI linewidth at half power ($W_{m50}$); $W_{m50}$ as function of heliocentric velocity; distribution of the logarithm of the integrated flux.
}
\label{hist_ADHI}
\end{figure*}

Figure~\ref{hist_ADHI} provides a graphic summary of aspects of the All Digital HI Catalog.  The redshift distribution in the top left panel contains peaks at 1,800~\kms\ and 5,000~\kms\ .  The two peaks are partially due to the development of the sample with the increasing capabilities of telescopes, with an early emphasis on the region within 3,000~\kms, and partially a reflection of the distribution of nearby structure.

In the top right panel it is seen that there is a tail to the line width distribution that extends to 1,000~\kms.  The galaxies that contribute to this tail are of sufficient interest that we will discuss them in a later publication.  It is seen in the lower left panel that galaxies with a full range of line widths are seen over a wide redshift range.  Within 3,000~\kms\ there is greater representation of narrow line width systems as a consequence of sample selection.    

\begin{figure}
\centering
\begin{tabular}{c}
\includegraphics[width=8.3cm]{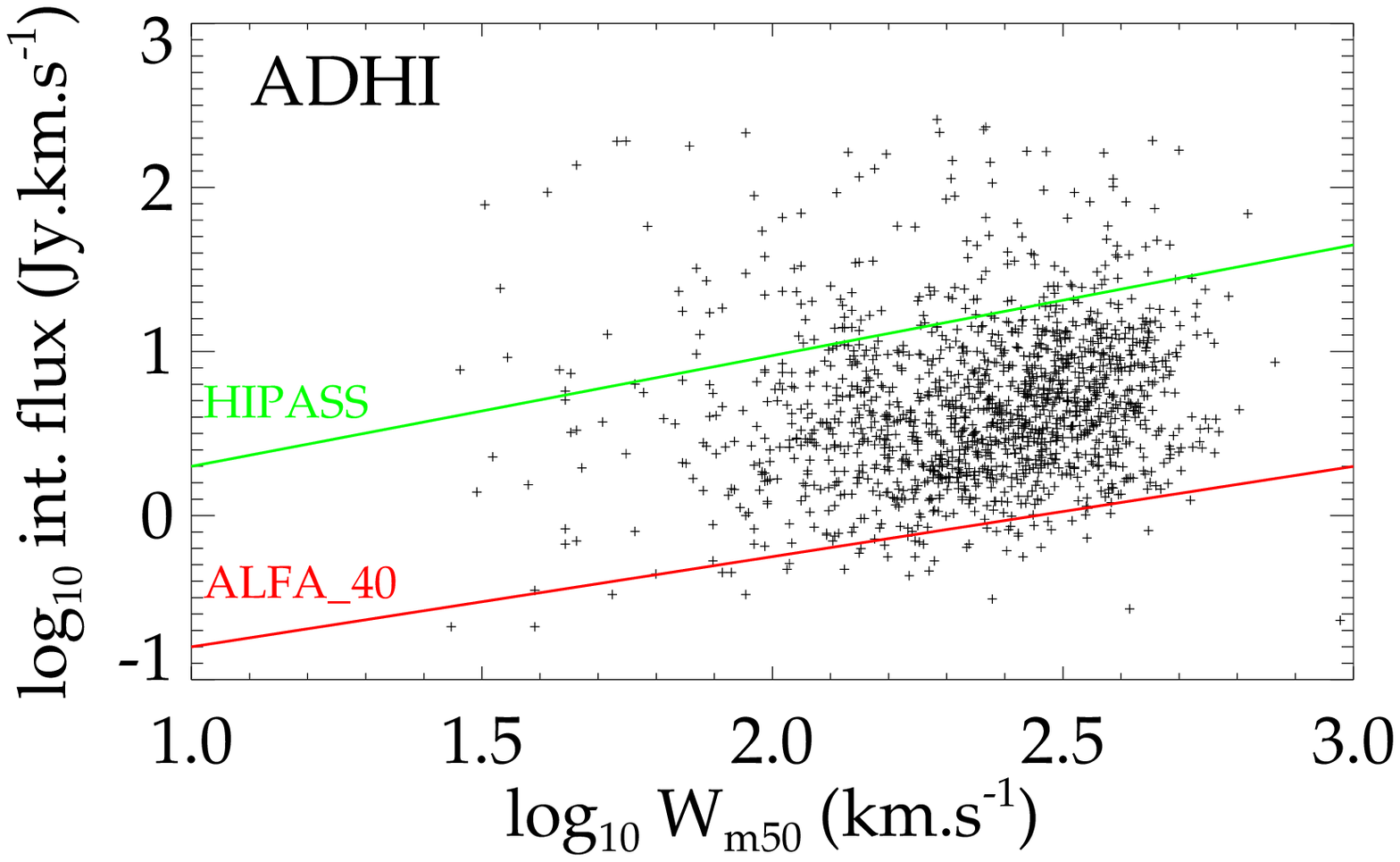} \\ 
\includegraphics[width=8.3cm]{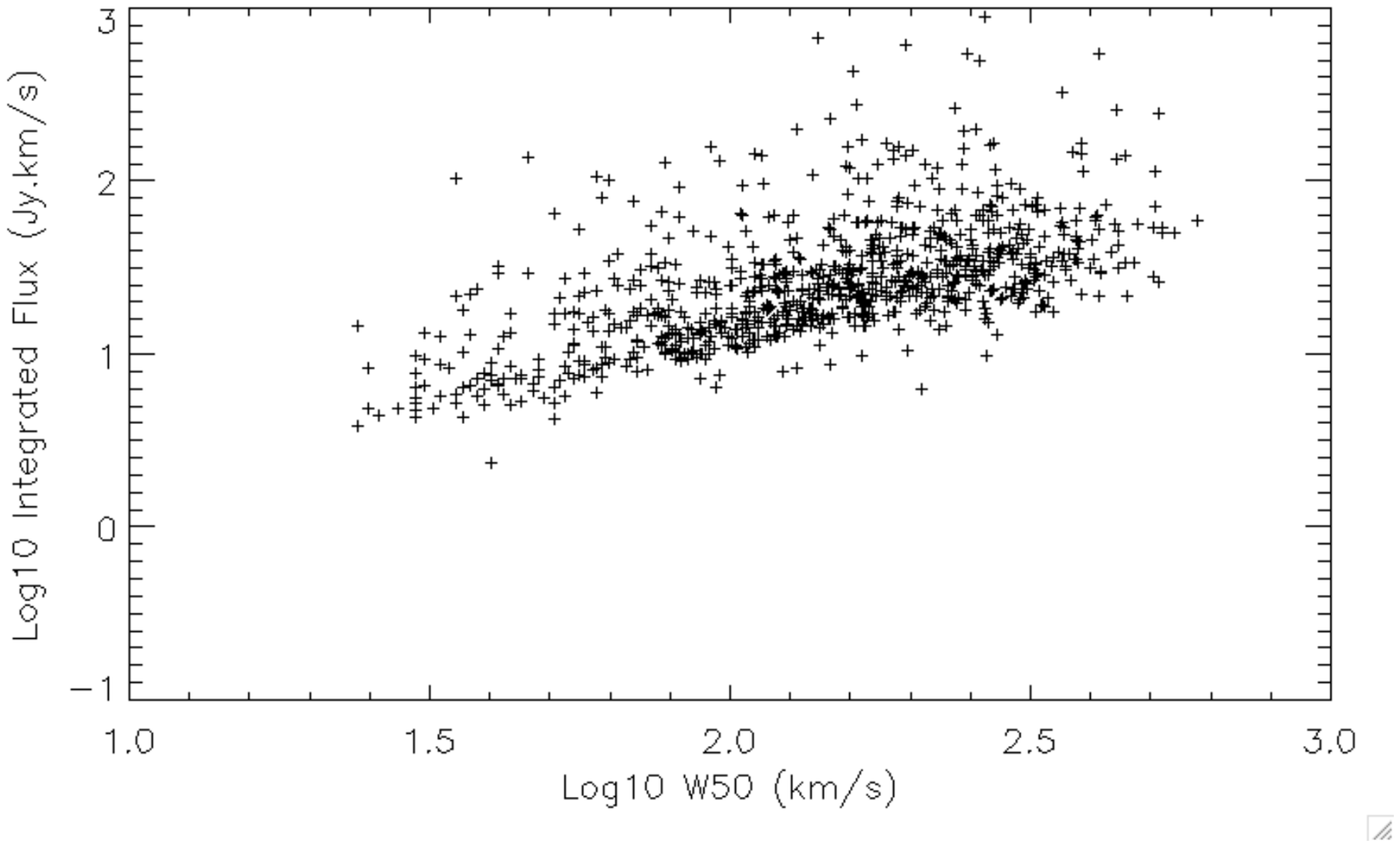}\\
\includegraphics[width=8.3cm]{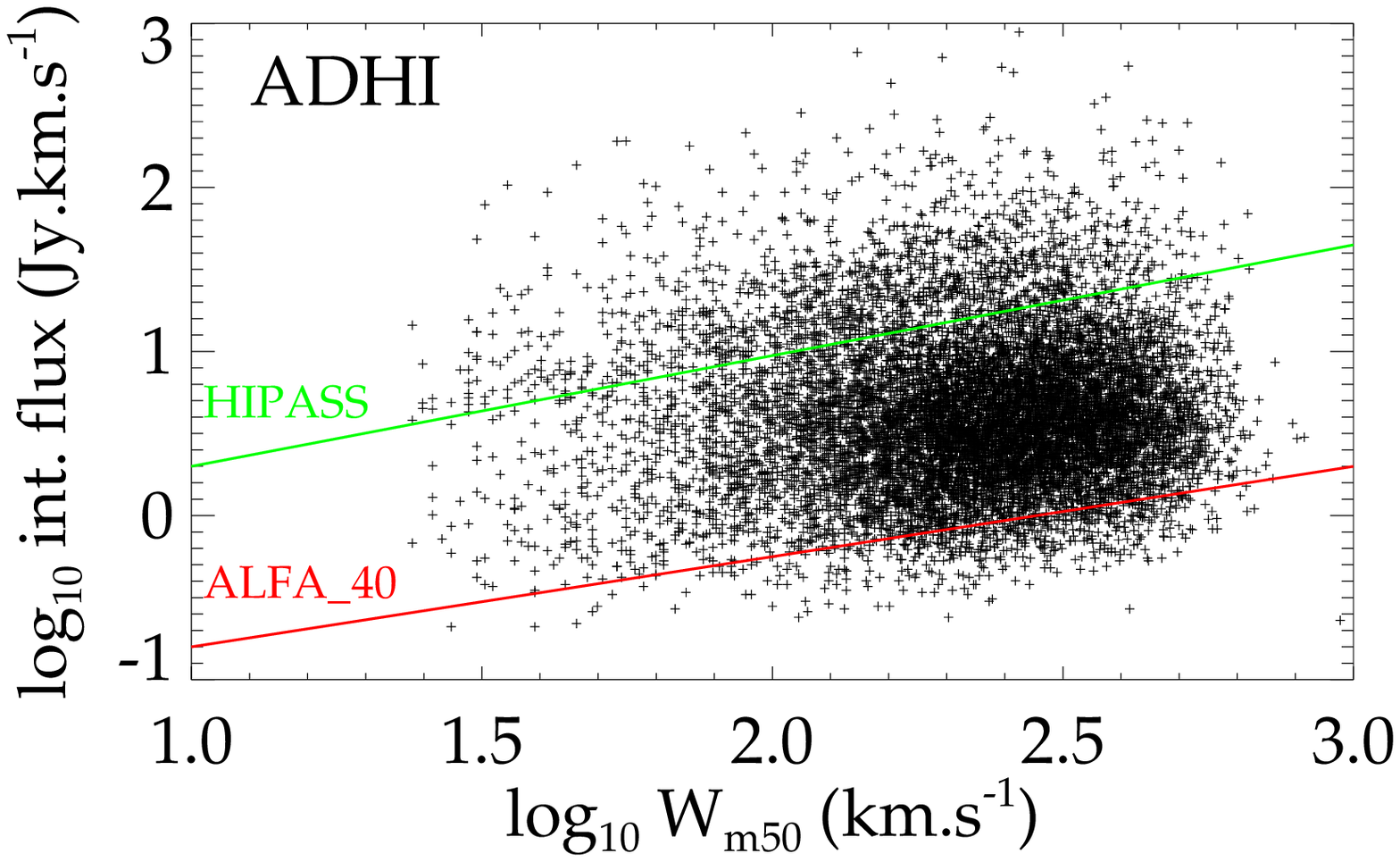} \\
\end{tabular}
\caption{
The distribution of sources detected with an adequate profile and an adequate signal-to-noise ratio:
(a) our GBT and Parkes observations, (b) 1000 brightest HIPASS profiles, (c) the All Digital HI Catalog with
overploted the detection limitation of HIPASS (green upper line) and ALFALFA 40\% (red lower line).
Our all sky catalog, which is mainly composed of targeted observations, doesn't show the detection bias trend of blind surveys. Blind surveys tends to not detect the low flux/high rotational velocity galaxies.
}
\label{detection}
\end{figure}

Since our full catalog is a compendium of many sources there are artifacts of the collection.  In Figure~\ref{detection} we plot the logarithm of the integrated flux vs. the logarithm of the linewidth.  The top panel shows that with our Cosmic Flows observations large line width targets are as equally well represented as small line width targets at the faint limit.  This situation arises because our integration times are established to be long enough to acquire adequate profiles case by case.  By contrast, in the middle panel it is seen that there is a strong dependence of the faint limit in the sample of the 1000 brightest galaxies in HIPASS \citep{2004AJ....128...16K} where integration times are fixed.  A similar trend though at a lower flux level is found in the ALFALFA survey after 40\% completion \citep{2010ApJ...723.1359M}.  Blind surveys with fixed integrations tend to miss low flux/high rotational velocity galaxies.   The bottom panel shows this trend is not severe for the ensemble of our All Digital HI catalog, since it is comprised at 85\% from targeted observations. For comparison we over-plot the detection limits for HIPASS (red upper line) and ALFALFA 40\% (green dashed line).

\begin{figure}
\includegraphics[width=8cm]{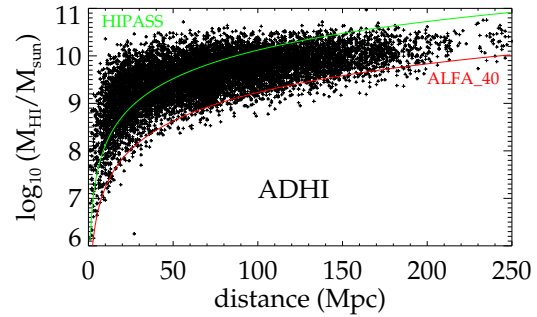} 
\caption{ The 11,051 accurate profiles plotted as $log(M_{HI}/M_{\sun})$ vs. distance in Mpc.
The adopted value to compute the distance here is $H_{0}$ = 70 km.s$^{-1}$.Mpc$^{-1}$.
The lower red solid line  is the ALFALFA 5 $\sigma$ sensitivity and the upper green line is the HIPASS 
detection limit.
}
\label{MHI_dist}
\end{figure}

Figure \ref{MHI_dist} shows the distribution of the detected masses with distances. 
The lower green solid line  is the ALFALFA 5 $\sigma$ sensitivity at 5.6 Jy.\kms and the upper red line is the HIPASS 
5 $\sigma$ detection limit at 0.372 Jy.\kms.

\begin{figure}
\includegraphics[width=8cm]{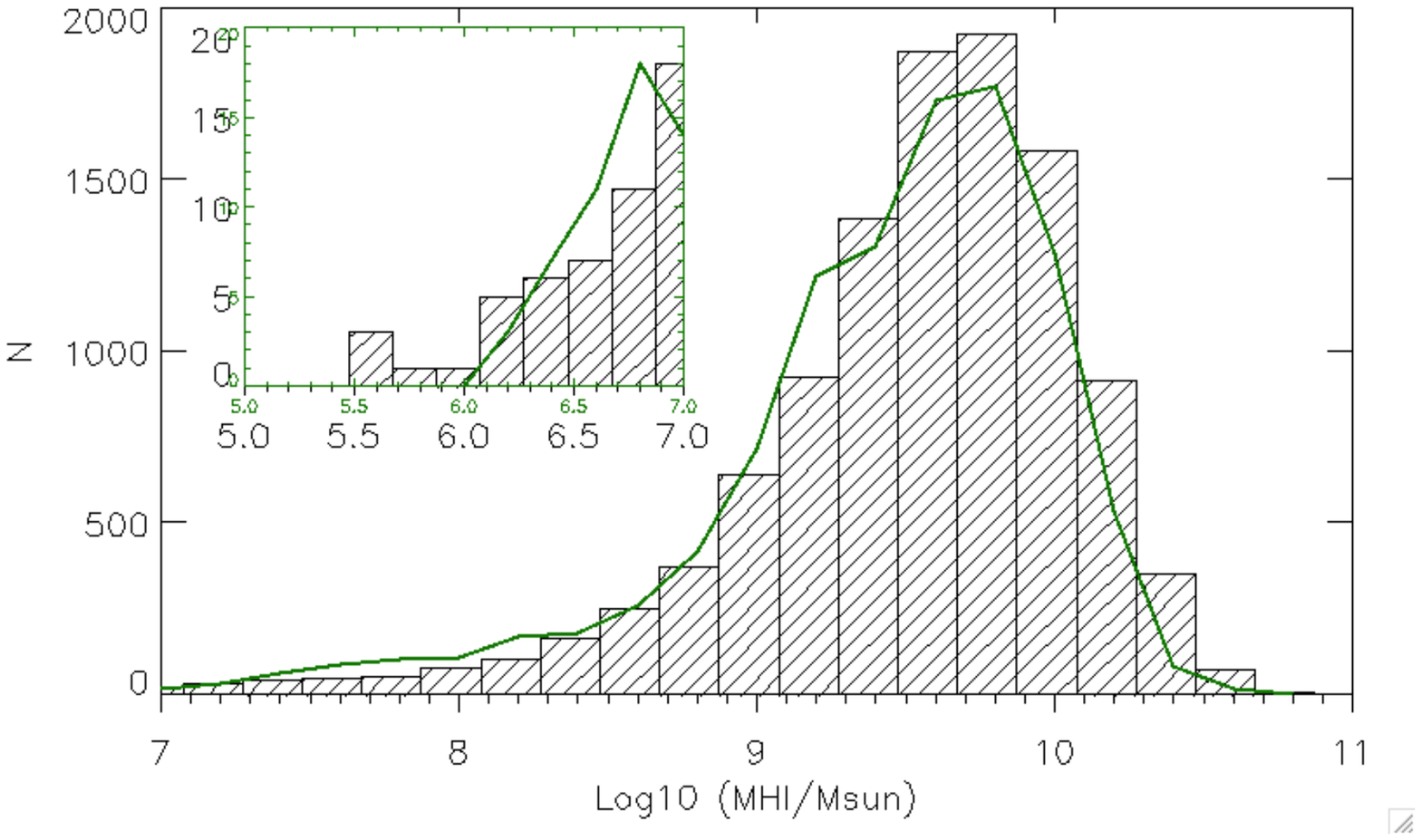} 
\caption{Histogram of the distribution of HI masses of 11,051 galaxies with good profiles, plotted as logarithm of the HI mass in solar units.
The results from 40\% ALFALFA are shown in red \citep{2010ApJ...723.1359M}. 
}
\label{HImass}
\end{figure}

HI mass is computed using the integrated flux and the redshift of the galaxies using:
\begin{equation}
M_{HI}=2.36 \times 10^{5} D^{2}_{Mpc} \times FI
\end{equation}
where $D_{Mpc}$ is the distance, and FI is the integrated flux in Jy.\kms.
For consistency in a comparison with ALFALFA velocities are converted to distance assuming $H_{0}$ = 70 km s$^{-1}$ Mpc$^{-1}$.  
Figure~\ref{HImass} shows the distribution of HI Mass in our all sky catalog. It is comparable with the ALFALFA 40\% HI Mass function from \citep{2010ApJ...723.1359M} displayed with the green line.  The cutoff at high HI mass is abrupt above $3 \times 10^{10}~M_{\sun}$.  The interesting objects in the high HI mass tail will be discussed in a separate publication along with the objects in the high line width tail.

\begin{table}
\caption{Radio-Telescopes in EDD (Extragalactic Distance Database)}
\label{tbl:tel}
\begin{tabular}{ccccr}
\hline
Telescope & Aperture & Beam & EDD & EDD  \\
                    &    meters              &   arcmin         &   Acronyms               &   spectra         \\
\hline
Arecibo     & 305       & ~3     & AOG-AOlf-ALFA & 7,898\\
Nan\c cay & 200x40 & 4x22 & Nanc  & 3,439\\
GBT           & 100       &  9    & GBT & 1,444 \\
GB300       & ~91        & 10     & GB300  & 1,059\\
Parkes       & ~64        & 14.5    & PAKS & 1,052\\
GB140       & ~43        & 21      & GB140 & 696 \\
Effelsberg & 100        & ~9       & Effs-Eff  & 235\\
\hline
\end{tabular}
\end{table}

\begin{table}
\caption{All Digital HI Catalog sources}
\label{tbl:sources}
\begin{tabular}{ll}
\hline
Code & Literature source\\
\hline
ksk2004 & Koribalski, Staveley-Smith, et al. 2004 \\
shg2005 & Springob, Haynes, Giovanelli, Kent 2005 \\
hkk2005 & Huchtmeier, Karachentsev, et al. 2005 \\
tmc2006 & Theureau, Martin, Cognard, et al. 2007 \\
ghk2007 & Giovanelli, Haynes, Kent, et al. 2007 \\
sgh2008 & Saintonge, Giovanelli, Haynes, et al. 2008 \\
kgh2008 & Kent, Giovanelli, Haynes, et al. 2008 \\
ctf2009 & Courtois, Tully, Fisher, et al. 2009 \\
ctm2010-ctk2010 & Courtois, et al. (this paper)\\
\hline
\end{tabular}
\end{table}

\section{Summary}

Over the years, many observers at the world's largest telescopes have acquired useful information about the neutral hydrogen properties of spiral galaxies.  Our primary concern is to measure distances by exploiting the correlation between galaxy HI profile line widths and luminosities, the TF relation.  Five samples have been defined: one to calibrate the TF relation with newly defined line width parameters and photometry, another to assure uniformity with the SNIa distance scale, and three more that provide all-sky coverage to different depths, with different densities, and with distinct selection criteria.  Observations with the Green Bank and the Parkes telescopes within the Cosmic Flows program have built upon the body of archival material to the degree that adequate HI profile information now exists for almost all the galaxies in our five samples.

Tabular and graphical information on the HI properties of galaxies, whether from new observations or from the archives, is gathered and made available at the Extragalactic Distance Database: http://edd.ifa.hawaii.edu, select the catalog `All Digital HI'.

\section*{acknowledgements}
New observations across the entire sky have been made possible by access to three fine radio telescopes.  We made early observations with the refurbished Arecibo Telescope and expect to add fresh material coming from the wide field multi-beam survey.  At the Green Bank Telescope our ongoing program Cosmic Flows has been awarded the status of a Large Program.    Observations of the deep southern sky began in 2009 with the Parkes Telescope in Australia.  The authors acknowledge the important support provided by GBT friends Franck Ghigo, Ronald J. Maddalena and  Toney Minter, GBT scheduling and direction team : Karen O'Neil, Jules Harnett and Carl Bignell, and all the operators who helped us conducting our 1000 hrs of observations: Dave Curry, Kevin Gum, Greg Monk, Dave Rose, Barry Sharp and Donna Stricklin. The authors also acknowledge the important support provided by CSIRO staff: Stacy Mader and Mark Calabretta for help in retrieving Parkes Archive material and for data flux calibration. Equally important to us has been access to archival material from the Cornell Digital HI Archive, the Nan\c{c}ay Radio Telescope HI profiles of Galaxies database, and the Australia Telescope online archive.  Although electronic archives are a great innovation, the low-tech information gathered in the {\it Pre Digital HI} catalog retains great value and we thank Cyrus Hall for his role in assembling that material.   We have made extensive use of NED, the NASA/IPAC Extragalactic Database operated by the Jet Propulsion Labratory, California Institute of Technology, and the HyperLeda database hosted at the Universit\'e Lyon 1.  Tully acknowledge support from the US National Science Foundation award AST-0908846. DM and IK were supported by the Russian Foundation for Basic Research
grants 08--02--00627, RUS-UKR 09--02--90414. 

\bibliographystyle{mn2e}

\label{lastpage}

\end{document}